\documentstyle[12pt]{article}

\newcommand{\bold}[1]{\mbox{\boldmath $#1$}}    %       bold symbol
\newcommand{\quart}{\mbox{${1\over4}$}}	%	1/4
\newcommand{\half}{\mbox{${1\over2}$}}		%	1/2
\newcommand{\threehalf}{\mbox{${3\over2}$}}	%	3/2
\newcommand{\ie}{{\em i.e.}}                    %       i.e.
\newcommand{\etal}{{\em et al.}}                %       et al.
\newcommand{\rme}{{\rm e}}      		%       roman e
\newcommand{\eps}{\epsilon}			%       epsilon
\newcommand{\ul}{\underline}			%       underline
\newcommand{\MeV}{{\rm MeV}}                    %       MeV
\newcommand{\fm}{{\rm fm}}                      %       fm
\newcommand{\del}{\partial}                     %       partial
\newcommand{\beq}{\begin{equation}}
\newcommand{\eeq}{\end{equation}}
\newcommand{\beqar}{\begin{eqnarray}}
\newcommand{\eeqar}{\end  {eqnarray}}
\renewcommand{\hbar}{}
\newcommand{\Bar}{\underline}
\newcommand{\zero}[1]{\raisebox{1.2ex}		%	put a 0 over a symbol
	{\hspace{0.2ex}\scriptsize{$\circ$}}\hspace{-1.1ex}{#1}}

\newcommand{\cappa}{{\mbox{\boldmath{\scriptsize{$\kappa$}}}}}
\newcommand{\Cappa}{{\mbox{\boldmath{{$\kappa$}}}}}
\newcommand{\kdotx}{{\bf k}\cdot{\bf x}}                %       k dot x
\newcommand{\k}{{\bf k}}                              %       bold k
\newcommand{\x}{{\bf x}}                              %       bold x
\newcommand{\Chi}{X}                                    %       Cap \chi
\newcommand{\argnull}{\mbox{$2\eta_0$+$\vartheta_0$}}
\newcommand{\argplus}{\mbox{$2\eta_0$+$\vartheta_0$+$\Delta$}}
\newcommand{\argminus}{\mbox{$2\eta_0$+$\vartheta_0$-$\Delta$}}
\begin{document}
%==============================================================================
\begin{titlepage}

\noindent{\sl Heavy Ion Physics}\hfill LBNL-42616\\[8ex]

\begin{center}
{\large {\bf Enhanced pion production in DCC dynamics$^*$}}\\[8ex]
{\sl J\o rgen Randrup}\\[1ex]

Nuclear Science Division, Lawrence Berkeley National Laboratory\\
University of California, Berkeley, California 94720
\\[6ex]
%\today
June 7, 1999\\[6ex]
{\sl Abstract:}\\
\end{center}

{\small\noindent
In order to elucidate the enhancement of pion production
that may occur during dynamical scenarios of interest
in connection with disoriented chiral condensates,
we study the evolution of boson modes
whose frequencies have a given arbitrary time dependence.
The quantum-field treatment yields expressions for
the time dependence of observables 
that depend only on the initial state and
specific state-independent enhancement coefficients
which can be obtained from the given evolution of the frequency.
It is shown how these coefficients
also can be obtained within an entirely classical framework
by judicious usage of the dependence of the resulting enhancement 
on the phase of the initial state.
Illustrative application is made for frequency evolutions
obtained from idealized simulations of high-energy heavy-ion collisions
and it is shown how the non-equilibrium evolution of the order parameter
may lead to significant enhancements in the final pion spectra.

\vfill
\noindent
$^*$This work was supported by the Director, Office of Energy Research,
Office of High Energy and Nuclear Physics,
Nuclear Physics Division of the U.S. Department of Energy
under Contract No.\ DE-AC03-76SF00098.\\
\noindent
}%      end small
\end{titlepage}

%==============================================================================
\section{Introduction}

Several years ago,
it was suggested that a high-energy collision
may lead to a transient approximate restoration of chiral symmetry
within the very excited collision volume and, furthermore,
the subsequent non-equilibrium relaxation of the chiral order parameter
might lead to long-wavelength isospin-directed oscillations
that were denoted as `disoriented chiral condensates' (DCC)
\cite{Anselm89,Anselm91,Blaizot92,RajagopalNPB399,Bjorken}.
Effective field theories present an important theoretical tool
for capturing the essential physics of this novel phenomenon,
as the underlying quantum-field theory
is too complicated to apply directly.
But even at the simpler level of an effective theory,
the scenarios encountered in heavy-ion physics are very challenging
because the dynamical evolution of the colliding system
requires that the field equations be solved in real time
and, moreover,
the system is often far away from equilibrium.

The most commonly employed effective model
is the linear $\sigma$ model \cite{sigma-model}.
It can be treated in an approximate manner
by means of the Hartree factorization technique,
which linearizes the field equation and yields
a description in terms of effectively independent quasiparticles
subject to a common time-dependent effective field
(which may be determined self-consistently
from the local quasiparticle distribution).
The evolution of the field,
whether self-consistent or prescribed,
then renders the effective quasiparticle mass time dependent.
Thus, the physics of the quasiparticle modes corresponds to that of
effectively free particles endowed with given time-dependent frequencies.
Instructive developments of particular relevance to
high-energy heavy-ion collisions have been made in recent years
\cite{CooperPRD50,CooperPRD51,BoyanovskyPRD51a,BoyanovskyPRD51b}.

The time dependence of the quasiparticle frequencies
may generally enhance the pion yield.
While the production of field quanta by a time-dependent field
is well known in electromagnetism,
the analogous phenomenon in harder to produce and explore
in strong-interaction physics.
An extreme example,
the temporary creation of a supercritical field,
may occur in a high-energy nucleus-nucleus collision
if the hot interaction zone cools down sufficiently rapidly
to generate a quench \cite{RajagopalNPB404,JR:PRL}.
If indeed such a quench occurs,
then spontaneous pion production will be possible,
in analogy with the electromagnetic scenario.
But an accurate description requires quantum-field theory,
as the vacuum fluctuations play an essential role as seeds of amplification.
More generally,
the production of particles by any time-dependent field
cannot be accurately described within the realm of standard classical physics.

This poses an important calculational obstacle,
as the most tractable numerical simulation methods
are basically classical in character and thus not adequate
for an accurate calculation of the enhancement phenomenon.
Thus, there is a need for further development
of the dynamical treatment of the effective field theory,
so that particle production by the time-dependent effective field
can be described.
The present study is intended to be a contribution towards this goal.
As a step in this direction,
we consider the simple situation when the spatial variation can be ignored,
as may be justified in the bulk of the collision zone.
With the restriction to idealized matter,
the effective quasiparticle masses are independent of position
and depend only on time and the discussion then simplifies significantly.

The present paper has several distinct parts.
First the needed quantum-field formalism is introduced
and the quasiparticle problem is brought onto diagonal form.
Subsequently,
a method originally conceived by Combescure \cite{Combescure}
is invoked to treat the evolution of each separate mode
and it is shown how the development of various observables
can be obtained simply in terms of specific state-independent coefficients.
This result brings out the equal role played by vacuum and statistical
fluctuations and elucidates how particle production can occur even
when the initial mode is unagitated.
Then we show how the same result for these coefficients
can be obtained entirely within a classical treatment
by means of a suitable decomposition of the dependence of the observables
on the phase of the initial state.
This result may prove useful for attempts to address the problem
within classical field theory which is practically more tractable.
Finally, the effect of the enhancement mechanism is illustrated
for idealized dynamical scenarios of interest in connection with the
DCC discussion.
It is shown that typical time dependencies of the effective pion mass
may lead to significant spectral enhancements.

%============================================================================
\section{Framework}

In the Hartree approximation,
the Hamiltonian density operator is given by
\beq\label{H}
\hat{\cal H}(\x,t)\ =\ \half
\left[\hat{\psi}(\x)^\dagger\hat{\psi}(\x) 
+ \nabla\hat{\phi}(\x)^\dagger\cdot \nabla\hat{\phi}(\x)
+ \hat{\phi}(\x)^\dagger \mu^2(t) \hat{\phi}(\x) \right]\ .
\eeq
%where the daggers are merely cosmetic,
%since the field operators are Hermitean.
%The total energy is then given by
%$E(t)=\int d^3\x \langle t|\hat{\cal H}(\x)|t\rangle$.
%
The field operators $\hat{\phi}(\x)$ and $\hat{\psi}(\x)$ in (\ref{H})
represent the local field strength and its time derivative,
respectively.
Since those are real, the associated field operators are Hermitean.
Therefore, in three spatial dimensions,
they can be Fourier expanded as follows,
\beqar
\label{phihat}
\hat{\phi}(\x)
&=& {1\over\sqrt{\Omega}}
\sum_\k {1\over\sqrt{2\zero{\omega}_\k}}
\left[ a_\k \rme^{i\kdotx}+a_\k^\dagger\rme^{-i\kdotx} \right]\ ,\\
\label{psihat}
\hat{\psi}(\x)
&=& {-i\over\sqrt{\Omega}}
\sum_\k \sqrt{\zero{\omega}_\k\over2}\
\left[ a_\k \rme^{i\kdotx}- a_\k^\dagger \rme^{-i\kdotx} \right]\ ,
\eeqar
where $\Omega$ is the volume.
The operator $a_\k^\dagger$ creates a free particle
having the wave vector $\bold k$.
Its frequency $\zero{\omega}_\k$
is given by $\zero{\omega}_k^2=k^2+m^2$,
where $m$ is its free mass,
and thus depends only on the magnitude of the wave vector, $k=|\bold{k}|$.
The free annihilation and creation operators
%$a_\k^{\phantom{\dagger}}$ and $a_\k^\dagger$ which 
satisfy the usual commutation relations for bosons,
$[a_{\k}^{\phantom{\dagger}},a_{\k'}^\dagger]=\delta_{\k\k'}$,
and it is elementary to verify the standard commutation relation
for the field operators,
$[\hat{\phi}(\x),\hat{\psi}(\x')]=i\delta(\x-\x')$.
Finally,
the total energy of the system is represented by the Hamiltonian operator
$\hat{H}(t)\ =\ \int d^3\x\ \hat{\cal H}(\x,t)$.

We may introduce the following ``pseudo-canonical'' variables,
\beqar
\label{Qk}
\hat{Q}_\k^{} &\equiv& 
{1\over\sqrt{2\zero{\omega}_k}}\ [a_\k^{} + a_{-\k}^\dagger]\
=\ \hat{Q}_{-\k}^\dagger\ ,\\
\label{Pk}
\hat{P}_\k^{} &\equiv& 
i\sqrt{\zero{\omega}_k\over2}\ [a_\k^\dagger - a_{-\k}^{}]\ =\
\hat{P}_{-\k}^\dagger\ .
\eeqar
They satisfy the usual canonical commutation relations,
$[\hat{Q}_\k,\hat{P}_{\k'}]=i\delta_{\k\k'}$
and the total Hamiltonian can be rewritten as
\beq
\hat{H}(t)\ =\ \sum_\k \half\left[\hat{P}_\k^\dagger\hat{P}_\k^{}
+\omega_k^2(t)\hat{Q}_\k^\dagger\hat{Q}_\k^{} \right]\ ,
\eeq
where $\omega_k^2=k^2+\mu^2(t)$.
However, the variables $\hat{Q}_\k$ and $\hat{P}_\k$
are {\em not} hermitean,
as is evident from eqs.\ (\ref{Qk}-\ref{Pk}),
and they are therefore not fully suitable for our treatment.

%----------------------------------------------------------------------------
%\subsection{Diagonalization}

In order to bring the Hamiltonian on a truly diagonal form
in terms of Hermitean canonical variables,
it is necessary to perform a unitary transformation
among any pair of modes having opposite momenta.
In fact, there is an entire family of such transformations
%(see eqs.\ (\ref{qk'}-\ref{pk'}))
and we find it notationally convenient to adopt the following particular one,
\beq\label{alpha}
\left\{
\begin{array}{ll}
\alpha_{\cappa=\k\phantom{-}} &\equiv\ {1\over\sqrt{2}} [a_\k + a_{-\k}]\\
\alpha_{\cappa=-\k} &\equiv\ {i\over\sqrt{2}} [a_\k - a_{-\k}]
\end{array}
\right.\ ,
\eeq
where we have arbitrarily assigned specific opposite signs for the two members
of each (degenerate) pair of opposite momentum labels $\k$ and $-\k$.
%for example according to the sign of $k_1$;
%any other labeling scheme will work as well. 
While the results will not depend on the particular convention,
the explicit expressions for the transformation
do depend on the assigned sign of the particular wave vector.
In order to emphasize these new diagonal modes
do not carry a directed momentum,
they are labeled by the greek index $\Cappa$ rather than $\k$.
Moreover, we will use $\zero{\omega}_{\kappa}=\zero{\omega}_k$
to denote the free frequency of the two degenerate diagonal modes
$\Cappa=\pm\k$.

These new operators satisfy the usual commutation relations,
$[\alpha_{\cappa}^{},\alpha^\dagger_{\cappa'}]=\delta_{\cappa\cappa'}$,
and the corresponding canonical operators are now Hermitean,
\beqar\label{qkhat}
\hat{q}_\cappa^{} &\equiv& {1\over\sqrt{2\hbar\zero{\omega}_\cappa}}\,
[\alpha_\cappa^{}+\alpha_\cappa^\dagger]\ 
=\ \hat{q}_\cappa^\dagger\ ,\\ \label{pkhat}
\hat{p}_\cappa^{} &\equiv& i\sqrt{\hbar\zero{\omega}_\cappa\over2}\ 
[\alpha_\cappa^\dagger-\alpha_\cappa^{}]\ =\ \hat{p}_\cappa^\dagger\ ,
\eeqar
with $[\hat{q}_\cappa,\hat{p}_{\cappa'}]=i\delta_{\cappa\cappa'}$.
We also note the form of the operator representing
the number of free quanta in the diagonal mode $\Cappa$, 
\beq
\hat\nu_\cappa\ =\ \alpha_\cappa^\dagger\alpha_\cappa^{}\ =\
{1\over2\zero{\omega}_\cappa^{}} \hat{p}_\cappa^2 
+ {\zero{\omega}_\cappa^{}\over2} \hat{q}_\cappa^2 -\half\ .
\eeq
The effective Hamiltonian 
is now fully separated into independent harmonic modes,
$\hat{H}(t)\ =\ \sum_\cappa  \hat{H}_\cappa(t)$,
where
\beq\label{Hpq}
\hat{H}_\cappa(t)\
%\half\zero{\omega}_\cappa\{\alpha_\cappa^\dagger,\alpha_\cappa^{}\}\
=\ \half\hat{p}_\cappa^2\
+\ \half{\omega}_\cappa^2(t)\ \hat{q}_\cappa^2\ ,	%=\
%(\hat{\nu}_\cappa+\half)\zero{\omega}_\cappa\ ,
\eeq
so each mode $\Cappa$ can be treated separately.

It is instructive to note that
the field operators (\ref{phihat}-\ref{psihat}) can be rewritten as follows,
\beqar\label{qk}
\hat{\phi}(\x) &=& \left({2\over\Omega}\right)^{1\over2}
\sum_{\cappa>0}	\left[ \hat{q}_\cappa\cos(\Cappa\cdot\x)
+\hat{q}_{-\cappa}\sin(\Cappa\cdot\x)\right]\ ,\\ \label{pk}
\hat{\psi}(\x) &=& \left({2\over\Omega}\right)^{1\over2}
\sum_{\cappa>0}	\left[ \hat{p}_\cappa\cos(\Cappa\cdot\x)
+\hat{p}_{-\cappa}\sin(\Cappa\cdot\x)\right]\ .
\eeqar
Consequently,
the coordinate operators $\hat{q}_\cappa$ represent the coefficients
in the trigonometric expansion of the field strength $\hat{\phi}(\x)$,
while the momentum operators $\hat{p}_\cappa$ represent the coefficients
in the corresponding expansion of its time derivative $\hat{\psi}(\x)$;
those with positive momentum index go with the {\em cosinus} terms
and the others do with the {\em sinus} terms
(this particular feature is a consequence of
the arbitrary enumeration convention
introduced with the definition in (\ref{alpha})).
Of course one could make the trigonometric expansion in many other ways.
This arbitrariness reflects the degeneracy of the modes with opposite
values of $\Cappa$ and corresponds to the freedom associated with a
rotation in the space spanned by those two modes.

%==============================================================================
\section{Treatment}

After the diagonalization of $\hat{H}(t)$
has been acheived as described above,
the evolution operator factorizes correspondingly,
\beq
\hat{U}(t,t_0)\ =\ \prod_\cappa \hat{U}_\cappa(t,t_0)\ .
\eeq
It is then possible to treat each mode $\Cappa$ separately,
using the method employed in ref.\ \cite{Combescure} for a single mode.
%(We therefore often omit the mode label $\Cappa$
%in the subsequent developments,
%which apply to any one particular mode.)
%----------------------------------------------------------------------------
%\subsection{Generating operators}
For each mode $\Cappa$, it is convenient to introduce the following
two-particle operators,
\beq\label{K}
\hat{K}_0 = \quart
\{\alpha_\cappa^{}\alpha_\cappa^\dagger
+\alpha_\cappa^\dagger\alpha_\cappa^{}\}\ ,\
\hat{K}_- = \half \alpha_\cappa^2\ ,\
\hat{K}_+ = \half {\alpha_\cappa^\dagger}^2\ .
\eeq
They satisfy the following SU(1,1) algebra \cite{Amado},
\beq\label{SU}
[\hat{K}_0,\hat{K}_\pm]=\pm \hat{K}_\pm\ ,\phantom{mm}\ 
[\hat{K}_-,\hat{K}_+]=2\hat{K}_0\ ,
\eeq
and we note that $\hat{K}_0^\dagger$=$\hat{K}_0^{}$ 
and $\hat{K}_-^\dagger$=$\hat{K}_+$.
%Moreover, we have
%$[2K_0,\alpha]$=$-\alpha$ and $[2K_0,\alpha^\dagger]$=$\alpha^\dagger$ as
%well as $[K_+,\alpha]$=$-\alpha^\dagger$ and $[K_-,\alpha^\dagger]$=$\alpha$.
%
The Hamiltonian for the mode $\Cappa$ can then be reexpressed,
\beq\label{HK}
\hat{H}_\cappa(t)\
=\ 2\omega_+(t)\hat{K}_0\ +\ \omega_-(t)[\hat{K}_+ + \hat{K}_-]\ ,
\eeq
where $\omega_\pm(t)\equiv(\omega_\kappa^2(t)\pm\zero{\omega}_\kappa^2)
/2\zero{\omega}_\kappa$ for each $\Cappa$.

%----------------------------------------------------------------------------
%\subsection{Squeezing}

It is now possible to define the following {\em queeze operators}
that depend linearly on a time-dependent complex parameter $\Gamma_\cappa(t)$,
\beq
\hat{A}_\cappa(t)\ \equiv\ 
\Gamma_\cappa \hat{K_+} - \Gamma_\cappa^* \hat{K_-}\ ,\
\ul{\hat A}_\cappa(t)\ \equiv\ 
\Gamma_\cappa \hat{K_+} + \Gamma_\cappa^* \hat{K_-}\ .
\eeq
We note that $\hat{A}_\cappa^\dagger$=$-\hat{A}_\cappa^{}$ 
and $\ul{\hat A}_\cappa^\dagger$=$\ul{\hat A}_\cappa^{}$.
If we write $\Gamma_\cappa=G_\cappa\exp(i\vartheta_\cappa)$ we may instead
use  $\gamma_\cappa=g_\cappa\exp(i\vartheta_\cappa)$ as the parameter,
where $g_\cappa=\tanh(G_\cappa)$ (ensuring $g_\cappa<1$),
and we have
\beq
\hat{A}\ =\ G [\hat{K}_+\rme^{i\vartheta} - \hat{K}_-\rme^{-i\vartheta}]\ ,\
\hat{\Bar{A}}\ =\ 
G [\hat{K}_+\rme^{i\vartheta} + \hat{K}_-\rme^{-i\vartheta}]\ ,
\eeq
for each mode $\Cappa$.
We may also introduce the squeezed annihilation operator
(satisfying
$\{\beta_\cappa^\dagger,\beta_{\cappa'}^{}\}=\delta_{\cappa\cappa'}$),
\beq\label{beta}
\beta_\cappa\ \equiv\ 
\rme^{\hat{A}_\cappa}\alpha_\cappa \rme^{-\hat{A}_\cappa}\ 
%=\ \alpha_\cappa\ \cosh G_\cappa\ 
%- \alpha_\cappa^\dagger\ \rme^{i\vartheta_\cappa}\sinh G_\cappa\
=\ [1-g_\cappa^2]^{-{1\over2}}
(\alpha_\cappa^{}-\gamma_\cappa \alpha_\cappa^\dagger)\ .
\eeq
%Then $\beta^\dagger=\cosh G(\alpha^\dagger-\gamma^* \alpha)$
%and $\exp(-A)\alpha\exp(A)=\cosh G(\alpha+\gamma \alpha^\dagger)$,
%of course.

%----------------------------------------------------------------------------
%\subsection{Evolution operator}

According to the method developed by Combescure \cite{Combescure},
the dynamics of each of the decoupled modes $\Cappa$
can be described in terms of a dimensionless complex parameter $\xi_\cappa(t)$
whose temporal evolution is governed by
\beq\label{xi}
\ddot{\xi}_\cappa\ +\ \omega_\kappa^2(t)\ \xi_\cappa\ =\ 0\ ,
\eeq
with suitable initial conditions (see later).
Once $\xi_\cappa(t)$ is known,
the following auxiliary quantities follow,
\beqar
z_\cappa(t)\ =\ \zero{\omega}_\kappa\xi_\cappa-i\dot{\xi}_\cappa\ &,&\ 
\varepsilon_\cappa=\arg(z_\cappa)\ ,\\
\bar{z}_\cappa(t)\ =\ \zero{\omega}_\kappa\xi_\cappa+i\dot{\xi}_\cappa\ &,&\
\bar{\varepsilon}_\cappa=\arg(\bar{z}_\cappa)\ .
\eeqar
It should be noted that $\bar{z}^{}_\cappa$ is {\em not} equal to
the complex conjugate $z^*_\cappa$.
The parameter $\gamma_\cappa$ is then given by
\beq\label{gammak}
\gamma_\cappa(t)\ = g_\cappa \rme^{i\vartheta_\cappa}\ 
\equiv\ {\bar{z}_\cappa \over z_\cappa} = 
{\zero{\omega}_\kappa\xi_\cappa+i\dot{\xi}_\cappa 
\over \zero{\omega}_\kappa\xi_\cappa-i\dot{\xi}_\cappa}\ ,
\eeq
so that its phase is given by 
$\vartheta_\cappa=\bar{\varepsilon}_\cappa-\varepsilon_\cappa$.
The time evolution operator for the mode is then \cite{Combescure}
\beq\label{U}
\hat{U}_\cappa(t,t_0)\ =\ \rme^{\hat{A}(\gamma_\cappa(t))}\
\rme^{-2i(\varepsilon_\cappa(t)-\varepsilon_\cappa(t_0))\hat{K}_0}\
\rme^{-\hat{A}(\gamma_\cappa(t_0))}\ .
\eeq

A number of simple cases having a constant frequency
are especially instructive.
With a view towards the later applications,
we reserve the notation $\zero{\omega}_\kappa$ for the frequency of free motion,
$\zero{\omega}_\kappa^2=\kappa^2+m^2$,
while $\omega_\kappa$ generally denotes the frequency
given by $\omega_\kappa^2=\kappa^2+\mu^2$
where $\mu^2(t)$ is the time-dependent effective mass squared.
For notational simplicity,
we shall usually drop the mode index $\Cappa$
and also often write $\omega_0$ for the free frequency.

%............................................................................
\subsection{Illustration: Free motion}
For free motion, $\mu^2=m^2>0$, we have
$\omega_\kappa^2=\zero{\omega}_\kappa^2\equiv \kappa^2+m^2$ at all
times.
A simple solution to (\ref{xi}) is then $\xi_\cappa(t)=\exp(i\zero{\omega}t)$,
corresponding to $\gamma_\cappa(t)=0$
and $\varepsilon_\cappa(t)=\zero{\omega}_\cappa t$.
In this case, $\hat{A}_\cappa$ vanishes
and the evolution operator becomes
$\hat{U}_\cappa=\exp(-2i\varepsilon \hat{K}_0)=\exp(-i\hat{H}_\cappa t)$,
which is recognized as the free evolution operator.
Furthermore,
the squeezed annihilation operator (\ref{beta}) is identical to the
free one, ${\beta}_\cappa={\alpha}_\cappa$.

We also note that a coherent state of the form (\ref{chi}),
$|\chi\rangle=\exp(\chi\alpha^\dagger-\chi^*\alpha)|0\rangle$,
satisfies the free Schr\"odinger evolution equation,
$i\del_t|t\rangle=\hat{H}|t\rangle$ 
(with $\hat{H}=\{\alpha^\dagger,\alpha\}\omega_0/2$),
{\em iff} the coefficient $\chi(t)$ satisfies $i\dot{\chi}=\omega_0\chi$
(which implies $\ddot{\chi}+\omega_0^2\chi=0$).

%............................................................................
\subsection{Illustration: Effective free motion}
\label{effective}
At finite temperature,
the application of the Hartree treatment 
to the linear $\sigma$ model leads to an approximate decoupling
into modes that are described by an effective mass, $\mu^2>0$
\cite{CooperPRD50,CooperPRD51,BoyanovskyPRD51a,BoyanovskyPRD51b,%
RajagopalNPB404,JR:PRD}.
Then $\omega_\kappa^2\equiv \kappa^2+\mu^2$ is positive and time independent
and we may again seek a solution to (\ref{xi})
that leads to a constant $\gamma_\cappa$.
Since we require $g_\cappa<1$ (recall $g_\cappa=\tanh G_\cappa$),
we must choose
\beq
\gamma_\cappa\ =\ g_\cappa\rme^{i\vartheta_\cappa}\ =\
-{\omega_\kappa-\zero{\omega}_\kappa \over
\omega_\kappa+\zero{\omega}_\kappa}
\ ,
\eeq
which is still real.
But it is negative when $\omega_k>\zero{\omega}_k$,
as is ordinarily the case in our applications
(because the temperature increases the mass),
and its phase is then $\vartheta_\cappa=\pi$.
For either sign, we have $\varepsilon_\cappa=\omega_\kappa t$,
corresponding to $\xi_\cappa=\exp(i\omega_\kappa t)$.
The evolution operator then takes on the expected form,
\beq
\hat{U}_\cappa(t,t_0)\ =\ 
\rme^{\hat{A}}\ \rme^{-i{\omega_\kappa\over2}
\{\alpha^\dagger_\cappa,\alpha^{}_\cappa\}t}\ \rme^{-\hat{A}}\ 
=\ \rme^{-i{\omega_\kappa\over2}\{\beta_\cappa^\dagger,\beta_\cappa^{}\}t}\ ,
\eeq
analogous to that of a free particle with the mass $\mu$,
where the squeezed operators are those introduced in eq.\ (\ref{beta}).
%$\beta$=$\exp(A)\alpha\exp(-A)$=$\cosh(G)(\alpha-\gamma\alpha^\dagger)$
%(which satisfy 
%$[\beta_\cappa^{},\beta_{\cappa'}^\dagger]=\delta_{\cappa\cappa'}$).

If the initial state is taken to be a coherent state of squeezed quanta
(see Eq.\ (\ref{chi0}) in the Appendix),
then the number of effective quanta is given by
$\tilde{\nu}=\langle\beta^\dagger\beta\rangle=|\tilde{\chi}|^2$,
whereas the number of real quanta depends on the phase $\tilde\eta$,
$\nu=\langle\alpha^\dagger\alpha\rangle=|\chi|^2
=(\omega_0/\omega)({\rm Re}\tilde{\chi})^2
+(\omega/\omega_0)({\rm Im}\tilde{\chi})^2$.
Moreover, the energy of the state is
\beq
E\ =\ \langle\hat{H}\rangle\ =\
%{\omega^2+\omega_0^2 \over 2\omega_0}(\chi^*\chi+\half)\ +\
%{\omega^2-\omega_0^2 \over 4\omega_0}((\chi^*)^2+\chi^2)\ =\
\omega_+(\chi^*\chi+\half)\ +\
\half\omega_-((\chi^*)^2+\chi^2)\ =\
(\nu+\half)\omega_+ + \nu\omega_-\cos2\eta\ ,
\eeq
as follows directly from (\ref{HK}),
and we see that $E$ ranges between $\nu\omega$ (obtained if $\chi$ is real)
and $\nu\omega_0$ (obtained if $\chi$ is imaginary).

%............................................................................
\subsection{Illustration: Supercritical motion}

In this case we have $\mu^2<0$ 
and we consider those (sufficiently soft) modes
for which $\Omega_\cappa^2\equiv-\kappa^2-\mu^2$ is positive.
We can achieve the simple initial value $\gamma_0=0$
by adopting the solution
$\xi\cappa=\exp(\Omega_\cappa t)-\gamma_+\exp(-\Omega_\cappa t)$,
where $\gamma_+ = (\omega_0+i\Omega_\cappa)/(\omega_0-i\Omega_\cappa)$.
This leads to
\beq
\gamma_\cappa(t)\ =\ 
\gamma_+ {\rme^{2\Omega_\cappa t}-1 \over \rme^{2\Omega_\cappa t} 
-\gamma_+^2}\ ,\,\,\,\
\varepsilon_\cappa(t)-\varepsilon_0\ =\
{\rm arg}\left(
{\rme^{2\Omega_\cappa t}-\gamma_+^2 \over 1-\gamma_+^2}\right)\ .
\eeq
Thus,
for large times the modulus $g_\kappa$ approaches unity and,
as we shall see in Sect.\ 5.1, %\ref{neg}: THE TEMPLATE IS NO GOOD FOR THIS!!
the particle number and related observables exhibit an exponential growth.

%============================================================================
\section{Evolution of observables}

In the general case when $\mu^2$ evolves in time,
we need to solve the equation (\ref{xi}) for each mode $\Cappa$,
which is a straightforward task,
once the initial conditions and $\mu^2(t)$ are given.
In scenarios encountered in high-energy nuclear collisions,
the initial state is often described in terms of
medium-modified quasi-particles with a suitable thermal effective mass.
If the systems expands and cools, as is expected be the case,
then $\mu^2(t)$ decreases rapidly
and usually approaches its free value in an oscillatory fashion.
In the most extreme cases,
it may even attain negative values for a while.
We wish to study such scenarios
by employing forms of $\mu^2(t)$ that have been
obtained in dynamical simulations with the linear $\sigma$ model.

Let us first consider the number of squeezed quanta in a given mode $\Cappa$,
as counted by the effective number operator
$\tilde{\nu}_\cappa\equiv\beta_\cappa^\dagger\beta_\cappa^{}$,
where we recall the definition of
the squeezed annihilation operator (\ref{beta}),
$\beta(t)=\exp(A)\alpha\exp(-A)$.
The expected number of squeezed quanta is then easy to calculate,
\beqar\nonumber
\tilde{\nu}(t) &=&
\langle t|\beta^\dagger\beta|t\rangle\ =\
\langle t_0|\rme^{A_0}\rme^{2i\varepsilon K_0}\rme^{-A}
\beta^\dagger\beta\
\rme^{A}\rme^{-2i\varepsilon K_0}\rme^{-A_0}|t_0\rangle\\ \label{nutilde}
&=& 
\langle t_0|\rme^{A_0}\rme^{2i\varepsilon K_0}
\alpha^\dagger\alpha\
\rme^{-2i\varepsilon K_0}\rme^{-A_0}|t_0\rangle\ =\
\langle t_0|\rme^{A_0} \alpha^\dagger\alpha\ \rme^{-A_0}|t_0\rangle\\ \nonumber
&=&
\langle t_0|\beta_0^\dagger\beta_0|t_0\rangle\ =\ \tilde{\nu}(t_0)\ ,
\eeqar
where we have used the fact that
the number operator $\alpha^\dagger\alpha$ commutes with $\hat{K}_0$.
The same calculation holds for any power $(\beta^\dagger\beta)^n$.
Thus the number distribution of squeezed quanta remains constant in time
and therefore it provides no information on the dynamics.

Using the relationship (\ref{beta}) between $\alpha$ and $\beta$,
we find
\beq
\beta_0^\dagger\beta_0\ =\ {1\over 1-g_0^2}
\left[ (1+g_0^2)\hat{\nu} +g_0^2 
-2\gamma_0 \hat{K}_+ -2\gamma_0^* \hat{K}_- \right]
\eeq
Therefore, if we start from a coherent state
of the form described in the Appendix,
\beq
|t_0\rangle=\rme^{\sum_\kappa
[\chi_\kappa^{}\alpha_\kappa^\dagger-\chi_\kappa^*\alpha_\kappa^{}]}
|0\rangle\ ,\
%\left({\hbar\omega_k \over \hbar^3c^3}\right)^{1\over2}C_k\ \rme^{-i\eta_k}\ ,
\eeq
where
$\chi_\kappa^{}=|\chi_\kappa^{}| \exp(-i\eta_\kappa)$,
we have
\beq
2\langle t_0|\hat{K}_0|t_0\rangle=|\chi|^2+\half\ ,\,\,\,\
2\langle t_0|K_\pm|t_0\rangle=|\chi|^2\rme^{\pm2i\eta}\ .
\eeq
Consequently, the initial number of physical quanta is
\beq
\nu_0\ \equiv\ \langle t_0|\hat{\nu}|t_0\rangle\
=\ \langle t_0|2\hat{K}_0|t_0\rangle\ =\ |\chi|^2\ ,
\eeq
while, with $\gamma_0=g_0\exp(i\vartheta_0)$,
the initial number of squeezed quanta is
\beq
\tilde{\nu}(t_0)\ =\
{1\over 1-g_0^2}
\left[ \left( 1+g_0^2 -2\cos(\vartheta_0+2\eta)\right)\nu_0 +g_0^2\right]\ .
\eeq

To extract interesting dynamical information,
it is useful to note that the subspace
spanned by the three rotation operators $\hat{K}_m$
is closed under the time evolution,
\beq\label{Km}
\hat{K}_m(t)\ \equiv\ \hat{U}^\dagger(t,t_0) \hat{K}_m \hat{U}(t,t_0)\
=\ \sum_{m'} Z_{mm'}(t)\ \hat{K}_{m'}\ .
\eeq
The complex time-dependent structure coefficients $Z_{mm'}$ can be determined
by elementary means.
Writing $Z_{mm'}=(1-g_0^2)^{-1}(1-g^2)^{-1}z_{mm'}$, we find
\beqar
{z}_{00}\hspace{0.35em} 
&=& (1+g_0^2)(1+g^2)\ -\ 4g_0g\cos\Delta\ ,\\
{z}_{0+}\hspace{0.15em}
&=& \left[-g_0(1+g^2) + (1+g_0^2)g\cos\Delta
+ i(1-g_0^2)g\sin\Delta\right]\rme^{i\vartheta_0} = {z}_{0-}^* ,\\
{z}_{++}
&=& \left[ -2g_0g + (1+g_0^2g^2)\cos\Delta + i(1-g_0^2g^2)\sin\Delta\right]
\rme^{-i(\vartheta-\vartheta_0)} = {z}_{--}^* ,\phantom{m}\\
{z}_{+0}\hspace{0.2em}
&=& 2\left[(1+g_0^2)g-g_0(1+g^2)\cos\Delta-ig_0(1-g^2)\sin\Delta\right]
\rme^{-i\vartheta} = {z}_{-0}^* ,\\
{z}_{-+}
&=& \left[ -2g_0g\ +\ (g_0^2+g^2)\cos\Delta - i(g_0^2-g^2)\sin\Delta\right]
\rme^{i(\vartheta+\vartheta_0)} = {z}_{+-}^* ,
\eeqar
where
$\Delta\equiv\varepsilon+\bar{\varepsilon}-\varepsilon_0-\bar{\varepsilon}_0$
and we recall $\gamma=g\exp(\vartheta)$.
We note that the diagonal coefficients start out from unity,
$Z_{mm}(t_0)=1$,
while the off-diagonal coefficients initially vanish.
Moreover, $Z_{00}$ can never become smaller than one,
\beq
Z_{00}\ \geq\ {(1+g_0^2)(1+g^2)-4g_0g \over (1-g_0^2)(1-g^2)}\
=\ {(1-g_0g)^2+(g-g_0)^2 \over (1-g_0g)^2-(g-g_0)^2}\ \geq 1\ .
\eeq
The evolution of any observable quantities
should not depend on the choice of initial values $\xi_0$ and $\dot{\xi}_0$,
because any choice leads to the same evolution operator $\hat{U}$.
This physical feature is reflected in the fact that the
above $Z$ coefficients are in fact all independent
of the initial conditions $\xi_0$ and $\dot{\xi}_0$.
Furthermore,
it is elementary to show that $\ddot{Z}_{00}(0)=\omega_-^2$,
which shows explicitly that the initial growth of $Z_{00}$
is independent of $\xi_0$ and $\dot{\xi}_0$
and depends only on the physical quantity $\omega_-$.
We finally note that the magnitude of each $Z$ coefficient
remains constant during time intervals where $\omega^2$ is unchanged.

That the fact that the functions $Z_{mm'}(t)$
are independent of the specific initial values of $\xi_0$ and $\dot{\xi}_0$
implies that they depend only on the magnitude $\kappa=|\Cappa|$,
since the equation (\ref{xi}) for $\xi_\cappa$
contains only $\omega_{\kappa}^2$.
Thus, in effect,
$Z_{mm'}(t)$ depends only on the initial energy $\omega_0$.

%............................................................................
\subsection{Expected particle number}

We now focus on the development of the expected number of free quanta
in each of the independent modes,
$\nu_\cappa(t)=\langle t|\alpha_\cappa^\dagger\alpha_\cappa^{}|t\rangle$
(keeping in mind that this quantity is distinct from
the number of quanta having a given momentum,
$n_\k(t)=\langle t|a_\k^\dagger a_\k^{}|t\rangle$).

Since the number operator can be expressed as
$\hat{\nu}\equiv\alpha_\cappa^\dagger\alpha_\cappa^{}=2K_0-{1\over2}$,
we find
\beq
\hat{\nu}(t)\ \equiv\ \hat{U}^\dagger\ \hat{\nu}\ \hat{U}\ =\
2Z_{00}(t)\hat{K}_0\
+\ 2Z_{0+}(t)\hat{K}_+\ +\ 2Z_{0-}(t)\hat{K}_-\ -\ \half\ .
\eeq
If we start from a coherent state of the standard form (\ref{chi}),
with $\chi$=$\sqrt{\nu_0}\exp(-i\eta)$,
we easily find
\beqar\label{N} \nonumber
\nu\ &=&\ Z_{00}(t)(\nu_0+\half)\ -\half\
+\ [Z_{0+}(t)\ \rme^{2i\eta_0}\ +\ Z_{0-}(t)\ \rme^{-2i\eta_0}]\nu_0\\
&=&(1-g_0^2)^{-1}(1-g^2)^{-1}\left\{
[(1+g_0^2)(1+g^2) - 4g_0g\cos\Delta](\nu_0+\half)\right.\\ \nonumber
&-&\left. 2\left[ g_0(1+g^2) \cos(\argnull) 
- g\cos(\argplus)
- g_0^2g\cos(\argminus) \right]\nu_0\right\}\ .
\eeqar
We note that if we start from vacuum
then $\nu_0=0$ and the resulting number of quanta, $\nu(t)$,
is given solely in terms of the coefficient $Z_{00}$,
$\nu=\half(Z_{00}-1)$,
which does not contain the initial phase $\eta_0$.
This is in accordance with our expectations,
since the vacuum has no definite phase.
In the present description this feature is automatically guaranteed
by the fact that the coherent-state parameter for the vacuum vanishes,
$\chi_{\rm vac}=0$,
and so the phase has no significance.

If we consider an ensemble of coherent states
for which the initial phase $\eta_0$ is random,
$\prec\rme^{i\eta_0}\succ$=0,
then the ensemble average of (\ref{N}) eliminates the terms containing $\eta_0$.
[We note that this phase average can be obtained exactly by averaging over
$N$ equally spaced angles $\eta_n=\eta_0+2\pi n/N$,
$n=1,\dots,N$, where $N$$\geq$2 and the value of $\eta_0$ is immaterial.]
The evolution of the particle number is then again given by the first term
in expression (\ref{N}),
which may be recast to exhibit the increase in the occupancy,
\beq
\prec\Delta\nu(t)\succ\ \equiv\ 
\prec\nu(t)\succ\ -\ \nu_0\ =\ (Z_{00}(t)-1)\ (\nu_0+\half)\ .
\eeq
That expression has a simple interpretation:
the field fluctuations,
which are initially proportional to $\nu_0+\half$,
are magnified by the factor $Z_{00}(t)$.
The total field fluctuations are represented by the operator $2K_0$
and contain both the vacuum fluctuations
(amounting to half a quantum for each mode $\Cappa$)
and the fluctuations associated with the real excitations of the mode
(represented by the number operator $\hat\nu$).
The above result may also be written on the form
\beq\label{XN}
\prec\nu(t)+\half\succ\ =\ X_N(t) \prec\nu(t_0)+\half\succ\ ,
\eeq
where we have introduced the
{\em number amplification coefficient} $X_N=Z_{00}$.
It follows from our earlier discussion that this amplifcation coefficient
depends only on the magnitude of the wave number,
which in turn implies that the original oppositely moving states $\pm\k$
are on the average affected
in the same manner by the time dependence of $\omega^2$.
However,
this is not true for the individual states
due to the phase-dependent terms in (\ref{N}).
We also note that the above expression
corresponds to the result derived by Boyanovsky \etal\
in the Hartree approximation \cite{BoyanovskyPRD51a,BoyanovskyPRD51b}.

%............................................................................
\subsection{Variance in particle number}

We turn now to the variance in the number of quanta in a given mode,
\beq
\sigma_\cappa^2(t)\ \equiv\ 
\langle t|\hat{\nu}_\cappa^2|t\rangle\ -\ \nu_\cappa(t)^2\ .
\eeq
Because of the subtraction,
we need only consider the commutator terms resulting from
bringing the square of the number operator on normal form.
We therefore find
\beqar
\sigma^2(t) &=&  \langle t_0
| Z_{00}^2 \hat{\nu} 
+ 4Z_{00}(Z_{0+}K_+ +Z_{0-}K_-) 
+ 4Z_{0+}Z_{0-}(\hat{\nu}+\half) |t_0\rangle\\
&=&
\left[Z_{00}^2+
2Z_{00}(Z_{0+}\rme^{2i\eta}+Z_{0-}\rme^{-2i\eta})+4Z_{0+}Z_{0-}\right]\nu_0\ 
+\ 2Z_{0+}Z_{0-}\ .\phantom{m}
\eeqar
If again we perform an ensemble average,
the terms with the phases $\eta$ disappear,
\beq
\prec\sigma^2(t)\succ\ =\
(Z_{00}^2+4Z_{0+}Z_{0-})\nu_0\ +\ 2Z_{0+}Z_{0-}\ =\
X_N^2\nu_0\ +\ 4|Z_{0+}|^2(\nu_0+\half)\ .
\eeq
Since $Z_{00}(t_0)=1$ and $Z_{0\pm}(t_0)=0$,
we initially have $\sigma^2(t_0)=\nu(t_0)$,
as is characteristic of the Poisson distribution implied by the coherent state.
But, except when the motion is free,
the value of $\sigma^2$ will generally differ from $\nu$ for $t>t_0$,
and hence the time-dependent state does {\em not} 
retain a standard coherent form
({\em cf.}\ Ref.\ \cite{Amado}).

%............................................................................
\subsection{Expected energy}

Finally,
we consider the evolution of the energy expectation value in a given mode,
\beq
E(t)\ =\ \langle t|\hat{H}|t\rangle\
=\ \langle t|2\omega_+\hat{K_0} + \omega_-(\hat{K}_++\hat{K}_-)|t\rangle\ ,
\eeq
where we recall that $\omega_\pm\equiv(\omega^2-\omega_0^2)/2\omega_0$.
Again, we may use the relation (\ref{Km}) to express the time-dependent
expectation values in terms of expectation values in the initial state,
\beqar\label{E}
E(t) &=&
\left[\omega_+Z_{00}\ +\ \omega_-{\rm Re}(Z_{+0})\right]
(\nu_0+\half)\\ \nonumber
&+& \left[2\omega_+ {\rm Re}(Z_{0+}\ \rme^{2i\eta_0})\
+\  \omega_-
{\rm Re}\left((Z_{++}+Z_{-+})\ \rme^{2i\eta_0}\right) \right]\nu_0\ .
\eeqar

Furthermore,
for an ensemble in which the expectation values
$\langle t_0|K_\pm|t_0\rangle=\nu_0\exp(\pm2i\eta_0)$ average to zero,
only the first line of (\ref{E}) contributes and we obtain
\beqar\nonumber
\prec E(t)\succ &=& 
\left\{\omega_+(t)Z_{00}(t)\ 
+\ \omega_-(t){\rm Re}\left(Z_{+0}(t)\right)\right\}
(\nu_0+\half)\
=\ X_E(t)\prec E(t_0)\succ\\
&=& (1-g_0^2)^{-1}(1-g^2)^{-1}	\left\{
\omega_+\left[(1+g_0^2)(1+g^2) - 4g_0g\cos\Delta\right]\right.\\ \nonumber
&+&\left.2\omega_-\left[ (1+g_0^2)g \cos\vartheta
- g_0\cos(\vartheta-\Delta)
- g_0g^2\cos(\vartheta+\Delta) \right]\right\}(\nu_0+\half)\ .
\eeqar
where $X_E=[\omega_+Z_{00}+\omega_-{\rm Re}(Z_{+0})]/\omega_+(t_0)$
is the {\em energy amplification coefficient}.
With $\eps\equiv E/\omega_+(t_0)-\half$,
the above result may be recast in a form analogous to (\ref{XN}),
\beq\label{XE}
\prec \eps(t)+\half\succ\ =\ X_E(t) \prec \eps(t_0)+\half\succ\ .
\eeq
% being the excitation energy in units of the free frequency,
%
We note that when the frequency has returned to its free value,
$\omega^2=\omega_0^2$ (as usually happens sooner or later),
then $\omega_-=0$ and $\omega_+=\omega_0$.
Thus the expression (\ref{E}) for the energy enhancement
becomes identical to expression (\ref{N}) for the enhancement of $\nu+\half$,
as one would expect since then $E=(\nu+\half)\omega_0$.

For motion with a {\em constant} positive value of $\omega$,
the quantity $\gamma$ remains constant in time,
$\gamma=\gamma_-=-(\omega-\omega_0)/(\omega+\omega_0)$,
and so we have
\beqar
X_N(t)\ =\ Z_{00}(t) &=& 
{\omega_+^2\over\omega^2}-{\omega_-^2\over\omega^2}\cos 2\omega t\ 
\geq\ {\omega_+^2-\omega_-^2 \over \omega^2} = 1\ ,\\
Z_{\pm0}(t) &=& -{\omega_+\omega_-\over\omega^2} [1-\cos 2\omega t]\  +\
i{\omega_-\over\omega}\sin2\omega t\ =\ -2Z_{0\pm}\ ,\\
Z_{++}(t) + Z_{-+}(t) &=& 
-{\omega_-^2\over\omega^2}\ +\ {\omega_+^2\over\omega^2}\cos2\omega t\
+\ i{\omega_-\over\omega} \sin2\omega t\ .
\eeqar
Then the time-dependent terms in (\ref{E}) cancel out, as one would expect,
but the resulting conserved energy depends on the initial phase $\eta_0$
if $\omega\neq\omega_0$,
\beq
E(t)\ =\ E(t_0)\ =\ (\nu_0+\half)\omega_+\ +\ \nu_0\ \omega_-\cos2\eta_0\ ,
\eeq
and obviously the energy amplification coefficient remains equal to unity,
$X_E(t)=1$.

%==============================================================================
\section{Classical treatment}

We have described above how the cranked harmonic oscillator
can be treated exactly with the tools of quantum field theory
and we have illustrated how the development of observables
can be obtained by means of
appropriate state-independent amplification coefficients
whose evolution depends only on the prescribed time dependence of the
frequency of the mode.
We now describe how the key results can be obtained alternatively
within the framework of classical dynamics,
since this may be practically useful
in the context of dynamical simulations.
We shall restrict the present discussion to scenarios
where the initial frequency is the free one,
$\omega^2(t<0)=\omega_0^2$.
%as is the case in most situations of practical interest.

%----------------------------------------------------------------------------
%\subsection{Phase-space dynamics}

The starting point is the expression (\ref{Hpq})
for the Hamilton operator $\hat H$
in terms of the canonical operators $\hat q$ and $\hat p$
which suggests considering the classical Hamiltonian function
\beq
{\cal H}(q,p)\ =\ \half p^2\ +\ \half\omega^2(t)\ q^2\ .
\eeq
The corresponding equations of motion are the familiar Hamilton equations,
\beq\label{qpEoM}
\dot{q}\ =\ p\ ,\phantom{mm}\ \dot{p}\ =\ -\omega^2(t) q\ .
\eeq

For a given initial quantum many-body state of standard coherent form,
$|\chi_0\rangle$,
we associate a specific classical phase point $(q_0,p_0)$, where
\beq\label{qp0}
q_0\ =\ \sqrt{(2\nu_0+1)/\omega_0}\ \cos\eta_0\ ,\phantom{mm}\
p_0\ =\ -\sqrt{(2\nu_0+1)\omega_0}\ \sin\eta_0\ .
\eeq
We recall that the complex coherent-state parameter is given on the form
$\chi_0=\sqrt{\nu_0}\exp(-i\eta_0)$,
with $\nu_0=|\chi_0|^2$ being the expected number of quanta
in the coherent state,
$\nu_0=\langle\chi_0|\alpha^\dagger\alpha|\chi_0\rangle$,
and $\omega_0$ being the initial (free) frequency which is positive.
We then solve the classical equations of motion (\ref{qpEoM}),
using $(q_0,p_0)$ as the initial condition
and with the same specified $\omega^2(t)$ %time-dependent frequency squared
as employed in the quantal calculation.

The time-dependent energy of the classical system, $\cal E$,
is then given by the value of the Hamiltonian function,
\beq
{\cal E}(t)\ =\
{\cal H}(q(t),p(t))\ =\ \half p(t)^2\ +\ \half\omega^2(t)q(t)^2\ .
\eeq
Furthermore, in order to extract number of quanta, we note that
\beq
\hat{\nu}+\half\ =\ 
{1\over 2\omega_0}\ \hat{p}^2\ +\ {\omega_0\over2}\ \hat{q}^2\ ,
\eeq
and we therefore define the corresponding classical observable,
\beq
{\cal N}(t)\ \equiv\ {1\over 2\omega_0}\ p(t)^2\ +\ {\omega_0\over2}\ q(t)^2\ .
\eeq

It is seen from eqs.\ (\ref{N}) and (\ref{E}) that the quantal result
for the time-dependent expectation values of particle number and energy,
$\nu$ and $E$,
depend in a very simple manner on the initial phase $\eta_0$,
each being a constant plus a second harmonic.
Therefore we may write
\beqar\label{NQeta}
\nu(t) &=& \prec\nu(t)\succ\ \,
+\ \,\delta\nu(t)\ \,\cos(2\eta_0-\delta_N)\ ,\\ \label{EQeta}
E(t) &=& \prec E(t)\succ\ 
+\ \delta E(t)\ \cos(2\eta_0-\delta_E)\ .
\eeqar
Since the basic equation of motion is the same,
when expressed in the canonical variables,
we expect a similar behavior of the classical result.
This is indeed borne out by our calculations
and we may therefore write the classical results on a similar form,
\beqar\label{NCeta}
{\cal N}(t) &=& \prec{\cal N}(t)\succ\
+\ \delta{\cal N}(t)\ \cos(2\eta_0-\delta_N')\ ,\\ \label{ECeta}
{\cal H}(t) &=& \prec {\cal H}(t)\succ\ 
+\ \,\delta {\cal H}(t)\ \cos(2\eta_0-\delta_E')\ ,
\eeqar
where the phases $\delta_N'$ and $\delta_E'$ governing the classical results
are found to be the same as those in the expressions (\ref{NQeta}-\ref{EQeta})
for the quantal case.

In the quantal case,
the coefficients in (\ref{NQeta}-\ref{EQeta})
are given in terms of the calculated $Z$ coefficients.
In the classical case,
the coefficients in (\ref{NCeta}-\ref{ECeta})
can be extracted by Fourier resolving an ensemble of dynamical results
with respect to the initial phase $\eta_0$.
Since only the zeroth and second harmonics are present,
it suffices to consider three separate dynamical histories
having their initial phases separated by $2\pi/3$.
This elementary analysis then yields the evolution
of the three parameters in each of the two expressions
(\ref{NCeta}) and (\ref{ECeta}).

The quantal result (\ref{NQeta}-\ref{EQeta}) may also be written as
\beqar\label{dXN}
\nu(t)+\half &=& (\nu_0+\half)X_N(t)\
+\ \nu_0\delta X_N(t) \cos(2\eta_0-\delta_N)\ ,\\ \label{dXE}
E(t)/\omega_0 &=& (\nu_0+\half)X_E(t)\ 
+\ \nu_0\delta X_E(t) \cos(2\eta_0-\delta_E)\ .
\eeqar
It is then seen that $\prec\nu\succ=(\nu_0+\half)X_N-\half$
and $\prec E\succ=(\nu_0+\half)\omega_0X_E$,
as we have found already.
Furthermore, $\delta\nu=\nu_0\delta X_N$
and $\delta E=\nu_0\omega_0\delta X_E$.
Thus the phase-independent terms are proportional to $\nu_0+\half$,
whereas the phase-dependent terms are proportional to $\nu_0$,
which is a reflection of the presence of vacuum fluctuations
in the quantal ground state.

By contrast,
in the classical treatment all terms scale in the same manner,
being proportional to $\nu_0+\half$.\footnote{The specific choice
of $\nu_0+\half$ for the action in the classical calculation
is not essential due to the universal scaling;
the only important thing is to always employ a {\em finite} value
for the action,
so that even the evolution of the quantal vacuum can be addressed.}
A particular classical calculation is therefore only
a quantitatively good approximation to the quantal result
when the difference between $\nu_0+\half$ and $\nu_0$ is immaterial,
\ie\ for large occupation numbers, $\nu_0\gg1$.
However, as the above analysis shows,
the described resolution of the phase dependence
has enabled us to separate the two terms in the classical treatment
and we can therefore readily define
the corresponding amplification coefficients,
\beqar
X_N'\ \equiv\ 
\phantom{n}{\prec{\cal N}\succ\over \nu_0+\half}\phantom{n}\ &,&\,
\delta X_N'\ \equiv\ 
\phantom{n}{\delta{\cal N}\over \nu_0+\half}\phantom{n}\ ,\\
X_E'\ \equiv\ {\prec{\cal H}\succ\over (\nu_0+\half)\omega_0}\ 
&,&\,
\delta E_N'\ \equiv\ {\delta{\cal H}\over(\nu_0+\half)\omega_0}\ .
\eeqar
Once these coefficients have been determined on the basis of
the classical calculation,
by analysis of three separate trajectories,
they can then be used in the expressions (\ref{dXN}) and (\ref{dXE})
to provide the evolution of the quantum expectation values
$\nu(t)$ and $E(t)$ for any initial phase $\eta_0$
and any value of the initial number of quanta $\nu_0$.
We emphasize that this method provides the exact quantum result
even though it has been obtained entirely by classical means.

%----------------------------------------------------------------------------
%\subsection{Test illustrations}

The above method provides an alternative means
for obtaining the values of the amplification coefficients $X_N$ and $X_E$
solely on the basis of classical mechanics.
In order to demonstrate the utility of the method,
we now consider a number of especially simple cases.

%............................................................................
\subsection{Test case: Constant negative $\omega^2$}
\label{neg}

We first consider the case when a free mode
is suddenly exposed to a supercritical field.
So we assume $\omega^2=\omega_0^2>0$ for $t<0$
and $\omega^2=-\Omega^2<0$ for $t>0$
and then easily find
\beq\label{Xneg}
X_N(t\geq0)\ =\ 1\ +\ 
2\left({\omega_0^2+\Omega^2 \over 2\omega_0\Omega}\right)^2 \sinh^2\Omega t\ .
\eeq
Thus if we start from vacuum, $\nu_0=0$,
the number of spontaneously created quanta grows approximately exponentially
while the supercritical value $\omega^2=-\Omega^2$ is maintained,
\beq
\nu_0=0:\phantom{m}\
\nu(t>0)\ =\ \half(X_N-1)\ =\
\left({\omega_0^2+\Omega^2 \over 2\omega_0\Omega}\right)^2
 \sinh^2\Omega t\ 
\sim\ \rme^{2\Omega t}\ .
\eeq

To obtain the number amplification coefficient
with the corresponding classical treatment,
we use the form $q=a\exp(\Omega t)+b\exp(-\Omega t)$
and readily find the initial phase point,
\beq
q_0=a+b = \sqrt{2\nu_0+1\over\omega_0}\ \cos\eta_0\ ,\
p_0=\Omega(a-b) = -\sqrt{(2\nu_0+1)\omega_0}\ \sin\eta_0\ ,
\eeq
from which the expansion coefficients are readily determined.
The time evolution is then
\beqar
q(t)=\,\,\,\sqrt{\phantom{|}\nu_0+\half\phantom{|}\over2\Omega}\,\,\left\{
\sqrt{\Omega\over\omega_0}[\rme^{\Omega t}+\rme^{-\Omega t}]\cos\eta_0\ +\
\sqrt{\omega_0\over\Omega}[\rme^{\Omega t}-\rme^{-\Omega t}]\sin\eta_0\right\}
\ ,&~&\\
p(t)=\sqrt{(\nu_0+\half)\Omega\over2}\left\{
\sqrt{\Omega\over\omega_0}[\rme^{\Omega t}-\rme^{-\Omega t}]\cos\eta_0\ -\
\sqrt{\omega_0\over\Omega}[\rme^{\Omega t}+\rme^{-\Omega t}]\sin\eta_0\right\}
\ .&~&
\eeqar
The time evolution of ${\cal N}\equiv p^2/2\omega_0+q^2\omega_0/2$ then follows
and we note that ${\cal N}_0=\nu_0+\half$.
If we average over the initial phase $\eta_0$ the mixed terms vanish,
leaving
\beq
X_N'\ \equiv\ {\prec{\cal N}\succ\over\nu_0+\half}\ =\
{1\over\nu_0+\half}\prec {1\over2\omega_0}p^2+{\omega_0\over2}q^2\succ\ =\
1+2\left({\omega_0^2+\Omega^2 \over 2\omega_0\Omega}\right)^2 \sinh^2\Omega t\ ,
\eeq
which is the exact same result (\ref{Xneg})
as obtained above with the quantal treatment.

\begin{figure}[htb]
\vspace*{-1.0cm}
%		 \insertplot{super-N.ps}
%...................
\vspace{100mm}
\includegraphics{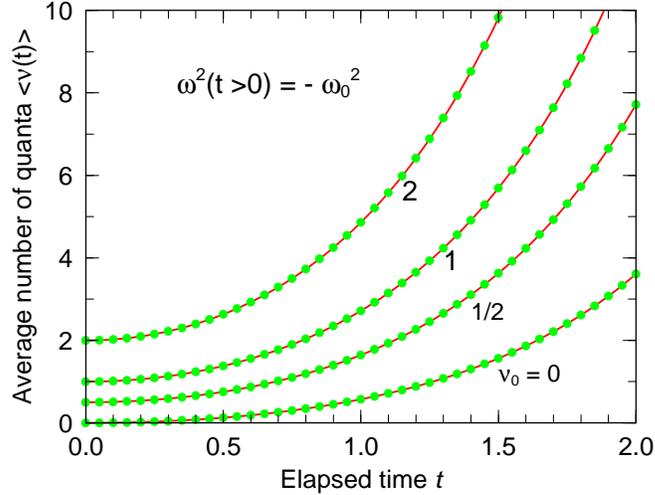}
%...................
\vspace*{-1.9cm}
\caption{Time dependence in supercritical scenario.}
{\small
The time dependence of the number of quanta in the system, $\nu(t)$,
for a simple supercritical scenario in which
$\omega(t<0)=\omega_0$ and $\omega^2(t>0)=-\omega_0^2$.
The number of free quanta $\nu$ as a function of the elapsed time $t$
since the sudden onset of supercriticality,
averaged over the initial phase $\eta_0$.
The values of the initial number of quanta $\nu_0$ are indicated.
Solid curves: quantum calculation;
solid dots: classical calculation.
Corresponding quantal and classical results are identical
to within the numerical tolerance.
}
\label{f:super-N}
\end{figure}

The resulting growth of the number of quanta
is illustrated in fig.\ \ref{f:super-N}
for the particular case where $\Omega=\omega_0$.
The results have been obtained by numerical solution
of the respective quantal or classical equations of motion,
as described earlier,
and the two treatments give the same result, $X_N'=X_N$,
as the above derivation demands.

%..............................................................................
%..............................................................................

\begin{figure}[htb]
\vspace*{-1.0cm}
%		 \insertplot{super-eta.ps}
%...................
\vspace{100mm}
\includegraphics{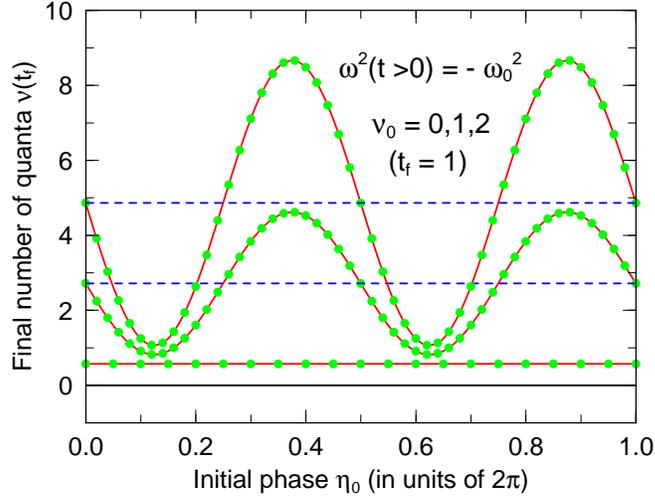}
%...................
\vspace*{-1.9cm}
\caption{Phase dependence in supercritical scenario.}
{\small
The number of free quanta $\nu(t_f)$ at the time $t_f=1$,
as a function of the initial phase $\eta_0$,
for the same supercritical scenario considered in fig.\ \ref{f:super-N}.
The initial number of quanta $\nu_0$ is indicated.
Solid curves: quantum calculation;
solid dots: classical calculation.
The corresponding phase-averaged result
is shown by the horizontal dashed line for each non-zero value of $\nu_0$.
Corresponding quantal and classical results are identical
to within the numerical tolerance.
}
\label{f:super-eta}
\end{figure}

For the evolution of individual states,
the agreement between the quantal and classical calculations
is illustrated in fig.\ \ref{f:super-eta}
which displays the number of quanta present
after $\omega^2$ has maintained its constant negative value $-\omega_0^2$
for a certain length of time, $\Delta t=1\ \fm/c$.
Since the phase $\eta_0$ is a cyclic variable,
the results are periodic in this parameter,
but the period is only $\pi$
since the phase dependence enters via the combination $2\eta_0$.
It should be noted that the two treatments are in full agreement as well
with regard to the phase-dependent results,
thus demonstrating that also the quantities governing the phase dependence
are identical for both amplitude and phase,
$\delta X_N'=\delta X_N$ and $\delta_N'=\delta_N$.

%..............................................................................
%..............................................................................

%............................................................................
\subsection{Test case: Constant positive $\omega^2$}
\label{pos}

When $\omega=\omega_0$ for $t<0$ and $\omega=const>0$ for $t>0$,
the quantal result is
\beq\label{XQplus}
X_N(t>0)\ =\ 
{\omega_+^2\over\omega^2}\ +\ {\omega_-^2\over\omega^2}\cos2\omega t\
=\ 1\ +\ {(\omega^2-\omega_0^2)^2 \over 2\omega^2\omega_0^2}\sin^2\omega t\ .
\eeq
To derive the classical result,
we use the form $q=a\cos\omega t + b\sin\omega t$
and readily find the initial phase point,
\beq
q_0=a=\sqrt{2\nu_0+1/\omega_0}\ \cos\eta_0\,\,\ ,\,\
p_0=\omega b=-\sqrt{2\nu_0+1\omega_0}\ \sin\eta_0\ .
\eeq
It is then elementary to obtain the amplification coefficient,
\beq\label{XCplus}
X_N'\ \equiv\
\prec {p^2+\omega_0^2 q^2\over(2\nu_0+1)\omega_0}\succ\
=\ \cos^2\omega t + 
\half({\omega^2\over\omega_0^2}+{\omega_0^2\over\omega^2})\sin^2\omega t\
=\ X_N\ .
%=\ 1+{(\omega^2-\omega_0^2)^2 \over 2\omega^2\omega_0^2}\sin^2\omega t\ .
\eeq
Thus, also in this case,
the quantal and classical expressions are identical.

We note that the number amplification coefficient oscillates
between its initial value $X_N(t=0)=1$ and its maximum value
$X_N^{\rm max}=(\omega^4+\omega_0^4)/2\omega^2\omega_0^2$,
and the temporal average is given by
$X_N^{\rm ave}=[(\omega^2+\omega_0^2)/2\omega\omega_0]^2$.
The amplitude increases with the difference between $\omega$ and $\omega_0$
and the frequency of the oscillation is equal to $2\omega$.
Therefore,
the resulting number of quanta in the system depends sensitively on
the duration of the perturbation.

%----------------------------------------------------------------------------
\subsection{Test case: Periodic positive $\omega^2$}
\label{osc}

\begin{figure}[htb]
\vspace*{-1.0cm}
%		 \insertplot{osc2-X.ps}
%...................
\vspace{100mm}
\includegraphics{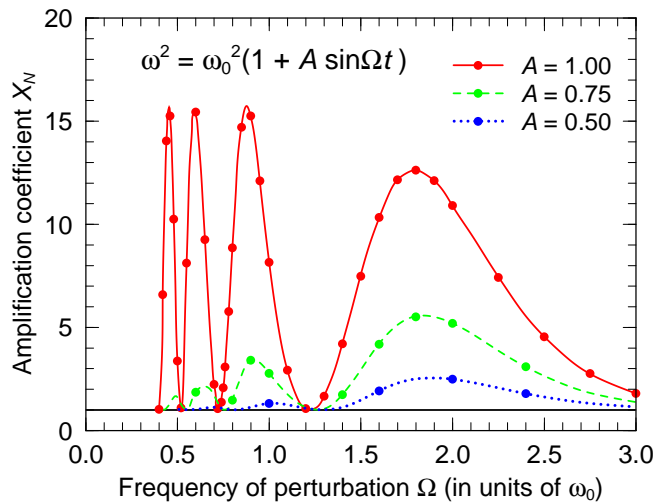}
%...................
\vspace*{-1.9cm}
\caption{Number enhancement in periodic scenario.}
{\small
The final number enhancement coefficient $X_N$
after the free oscillator has been subjected
to two periods of a harmonic perturbation,
$\omega^2=\omega_0^2[1+A\sin\Omega t]$,
as a function of the perturbation frequency $\Omega$ in units of $\omega_0$;
three values of the amplitude $A$ have been considered,
$A=0.50, 0.75, 1.00$.
The curves have been obtained by means of the quantum calculation,
$X_N^{\rm Qu}$,
while the solid dots result from the corresponding classical calculation,
$X_N^{\rm Cl}$.
}
\label{f:osc2-X}
\end{figure}

Finally,
we consider the special case of a harmonic perturbation of free motion,
\beq
\omega^2(t)\ =\ \omega_0^2\ [1+A\sin\Omega t]\ .
\eeq
The equation of motion then has the Mathieu form \cite{Mathieu}
and so we expect a resonant behavior 
as a function of the perturbation frequency $\Omega$.
This is illustrated in fig.\ \ref{f:osc2-X}
for the number enhancement coefficient $X_N$
resulting from subjecting the system to
two periods of the harmonic perturbation.
For small amplitudes and many periods these are given by
$\Omega=2\omega_0/n$, where $n=1,2,3,\dots$ is a natural number.
The resonance frequencies shift downwards
when the perturbation amplitude $A$ is increased.
Moreover,
it is of course not possible to accumulate much amplification
during only two periods unless the amplitude is large.
While the low frequency region, $\Omega<\omega_0$,
is characterized by rapid oscillations,
with an average $X_N$ value of around eight when $A$ is unity,
the higher region exhibits one broad bump centered somewhat below $2\omega_0$.
This latter feature implies that a given perturbation frequency $\Omega$
is especially effective in enhancing those modes
that have about half that frequency, $\omega_\kappa\approx\half\Omega$.
It should be noted that the induced enhancement
obtained with the classical treatment agrees with the quantal results.

%..............................................................................
%..............................................................................

\begin{figure}[htb]
\vspace*{-1.0cm}
%		 \insertplot{osc2-eta.ps}
%...................
\vspace{100mm}
\includegraphics{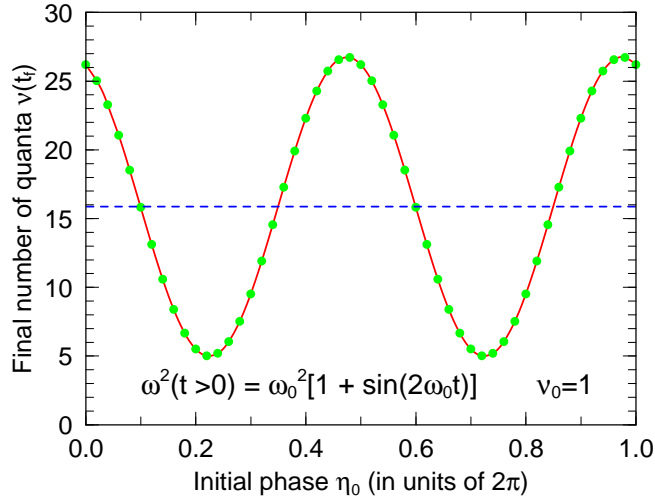}
%...................
\vspace*{-1.9cm}
\caption{Phase dependence in periodic scenario.}
{\small
The final number of free quanta $\nu_f$
as a function of the initial phase $\eta_0$,
after exposing the free oscillator to two periods of a harmonic perturbation,
$\omega^2=\omega_0^2[1+\sin2\omega_0t]$
(corresponding to an amplitude $A$=1 and a frequency $\Omega=2\omega_0$);
the results are shown for unit initial occupancy, $\nu_0$=1.
Solid curves: quantum calculation;
solid dots: classical calculation.
The phase-averaged results are shown by
the dashed horizontal line for each $\nu_0$
(it is the same for {\em Qu} and {\em Cl} to within the line width).
}
\label{f:osc2-eta}
\end{figure}

The dependence on the initial phase $\eta_0$
is shown in fig.\ \ref{f:osc2-eta} for the large amplitude, $A=1$,
in the case when one initial quantum is  present, $\nu_0=1$.
The behavior is qualitatively similar to
what was shown in fig.\ \ref{f:super-eta},
with the classical calculation tracking the quantal behavior perfectly.
The average value of the final occupancy $\nu_f$ is about 16,
in accordance with fig.\ \ref{f:osc2-X}
where the amplification factor $X(\Omega=2\omega_0)$ is seen to be about 11
(indeed, $\nu_f=(\nu_0+\half)X-\half\approx\threehalf\cdot11-\half=16$).

%..............................................................................
%..............................................................................

%............................................................................
\subsection{Semi-classical treatment}
\label{sc}

In many numerical studies
the field operator is replaced by a classical field,
$\hat{\phi}(\x)\to\phi(\x)$,
and the equation of motion can then be solved by standard means
(see refs.\ \cite{JR:PRL,AHW} 
for examples involving the linear $\sigma$ model).
In this type of approach,
the field strength is taken to represent the degree of agitation,
without allowing for any contribution from the vacuum.
Thus the number of quanta present in a given mode $\Cappa$
is taken to be proportional to the square of the corresponding field amplitude,
$\nu_\cappa\sim|\phi_\cappa|^2$.
This corresponds to using $\nu_0$ rather than $\nu_0+\half$
as the action in the classical phase-space dynamics (see eq.\ (\ref{qp0})).
Thus,	%As a consequence,
the quantum vacuum would be mapped into the ground state
of the classical oscillator,
$|0\rangle\to(q=0,p=0)$,
where it would then remain for any given form of $\omega^2(t)$,
thus precluding the spontaneous creation of particles from the vacuum.

Moreover,
the final phase-averaged action would become $\nu_f^{\rm sc}=\nu_0X_N$
which is always smaller than the quantal result,
$\nu_f^{\rm qu}=(\nu_0+\half)X_N-\half$,
since $X_N>1$.
While the two results are practically similar for large actions,
$\nu_0\gg1$,
the typical situations encountered involve occupation numbers well below unity.
In such cases the semi-classical treatment then
leads to a significant underestimate of the degree of particle creation
induced by the imposed external field.
We shall return to this important point
in the discussion of the quench scenario in Sect.\ 6.3 %\ref{quench}
and a specific example is shown in fig.\ \ref{f:400}.

The above described treatment makes it possible to circumvent
these inherent shortcomings of the classical treatment
and extract results from the classical calculation
that are quantitatively accurate for any degree of occupation.

%==============================================================================
\section{Application to DCC dynamics}

We here consider forms of $\omega^2(t)$ are especially relevant to scenarios
of interest in the context of disoriented chiral condensates.
They correspond to the conditions prevailing in the interior
of initially hot but rapidly expanding matter.

%----------------------------------------------------------------------------
\subsection{Initialization}

In the discussion of disoriented chiral condensates,
it is usually assumed that the early part of the nuclear collision
generates a very excited volume
within which chiral symmetry is at least partially restored.
The isospin degrees of freedom are then described in terms of
effective quasipions characterized by a medium-modified dispersion relation,
$\omega^2=k^2+\mu^2(t)$.
The effective mass depends both on the magnitude of the
chiral order parameter and the degree of agitation
which both evolve dynamically.
The initial degree of agitation can conveniently be characterized
by an equivalent temperature $T_0$
which may significantly exceed the free pion mass $m_\pi$.
The corresponding effective mass $\mu_0$ is then also significantly larger
than $m_\pi$.
In our applications,
we shall employ estimates obtained with the linear $\sigma$ model
which has been the most popular tool for $DCC$ studies.

After its preparation in a state of high excitation,
the system is expected to expand rapidly
and the corresponding sudden drop in the effective mass
may lead to quench in which the effective field has become supercritical,
$\mu^2<0$.
In order to generate such a quench,
it is necessary to first achieve approximate restoration of chiral symmetry
and this in turn requires a fairly high initial agitation,
corresponding to an equivalent temperature of several hundred MeV
\cite{JR:PRL}.
To illustrate this we shall employ $T_0=400\ \MeV$
which ensures that the initial order parameter is small,
$\sigma_0<10\ \MeV$;
the corresponding effective pion mass is taken as $\mu_0=560\ \MeV/c^2$.
(We recall that in vacuum the order parameter is equal to the pion decay
constant, $\sigma_0=f_\pi\approx92\ \MeV$.)
Alternatively,
in order to explore the possibility that such high initial excitations
are in fact not attained, we shall also consider a more moderate value,
$T_0=200\ \MeV$, for which the order parameter is still fairly large,
$\sigma\approx60\ \MeV$;
the corresponding effective pion mass is taken as $\mu_0=240\ \MeV/c^2$.
The specific parameter values are based on 
the semi-classical studies in ref.\ \cite{JR:PRD}
which lead to results that do not differ essentially
to those obtained by other methods
(such as the recently developed optimized perturbation theory \cite{Hatsuda}).

%..............................................................................
\begin{figure}[htb]
%\vspace*{-1.0cm}
%\vspace*{-2.9cm}
%		 \insertplot{f0.ps}
\vspace{80mm}
\includegraphics{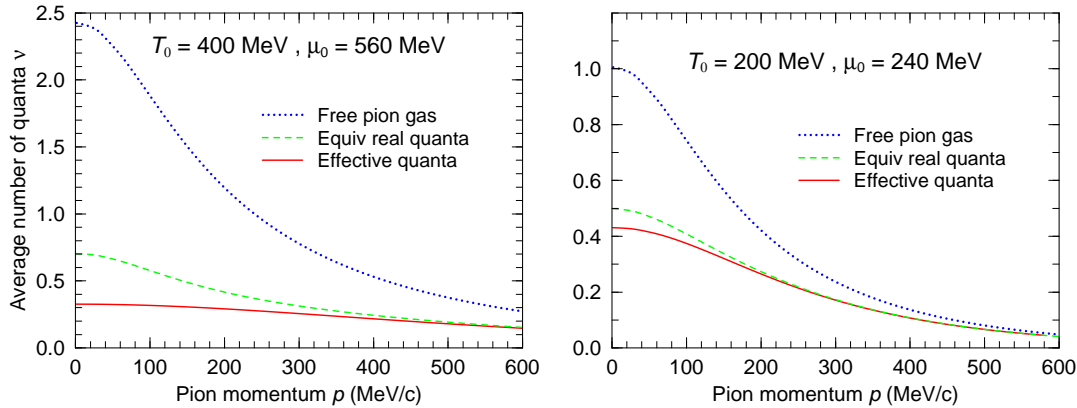}
\vspace*{-1.7cm}
\caption{Initial number of quanta.}
{\small
The solid curve shows the initial number of effective quanta 
$\tilde{\nu}_0$
in modes for which the momentum has the magnitude $p$,
under the assumption of thermal equilibrium at the temperature $T_0$
and using a corresponding effective pion mass $\mu_0$
as given by the linear $\sigma$ model \cite{JR:PRD};
the dispersion relation for initial effective pion mode is then
$\omega_i^2=p^2+\mu_0^2$.
The dashed curve shows the corresponding number of real quanta,
$\nu_0=\tilde{\nu}_0\omega_+/\omega_i$.
The thermal occupation number for a free pion gas
($\mu_0=m_\pi$) is also shown (dotted curve).
}
\label{f:f0}
\end{figure}
%..............................................................................

In fig.\ \ref{f:f0} we illustrate the relationship between the various
occupation numbers of interest.
We assume that the initial state is of the coherent form (\ref{chi0})
defined in terms of the effective (or squeezed) quasipion states.
In the present discussion,
we are interested only in quantities averaged over the initial thermal
ensemble and then the results depend only on the magnitude
of the state index, $\kappa=|\Cappa|=|\k|=k$,
which we shall therefore replace by the magnitude of the pion momentum, $p$.
(The momentum is a constant of motion in the present scenario where all the
dynamics arises from the time dependence of $\mu^2$.)
The initial energy $\omega_i$ of a given mode
is then given by the dispersion relation
$\omega_i^2=p^2+\mu_0^2$.
When expressed in terms of the quasipions,
the initial Hamiltonian operator is given by
$\hat{H}(t_0)=\half\omega_i\{\beta^\dagger,\beta^{}\}$,
as we have seen,
and the thermal average of the occupation number is therefore
of the usual Bose-Einstein form,
\beq
\tilde{\nu}_0\ \equiv\
\prec\langle\tilde{\chi}_0|\beta^\dagger\beta^{}|\tilde{\chi}_0\rangle\succ
=\ [\exp({\omega_i\over T_0})-1]^{-1}\ .
\eeq
This quantity is always well below unity
(in fact it never exceeds about one half at any temperature \cite{JR:PRD})
and it decreases towards a constant value at high temperatures
because the effective pion mass then grows in proportion to $T_0$).

The number of real pion quanta is represented by the number operator
$\alpha^\dagger\alpha^{}$ and it follows from our earlier analysis
that we have
\beq
\nu_0\ \equiv\
\prec\langle\tilde{\chi}_0|\alpha^\dagger\alpha^{}|\tilde{\chi}_0\rangle\succ\
=\ {\omega_+\over\omega_i}\ \tilde{\nu}_0\ \geq \tilde{\nu}_0\ ,
\eeq
since $\omega_+=\half(\omega_i^2+\omega_0^2)/\omega_0$.
This quantity is always larger than $\tilde{\nu}_0$,
especially for the softest modes where the modification of the mass
has the greatest influence.
It follows that the form of $\nu_0(p)$ cannot be approximated
by a free pion gas at a suitably adjusted effective temperature temperature.
Finally,
it is important to realize that the corresponding number
for a non-interacting gas of free pions at the given temperature $T_0$
would be significantly higher,
due to the (unjustified) use of the free mass $m_\pi$.

%----------------------------------------------------------------------------
\subsection{Evolution of the mass}

In order to emulate the conditions that may be experienced
inside the expanding collision zone,
we employ two different forms for $\mu^2(t)$
corresponding to two opposite idealized expansion scenarios.

\begin{figure}[htb]
%\vspace*{-1.0cm}
%\vspace*{-2.9cm}
%		 \insertplot{mu2.ps}!!
\vspace{80mm}
\includegraphics{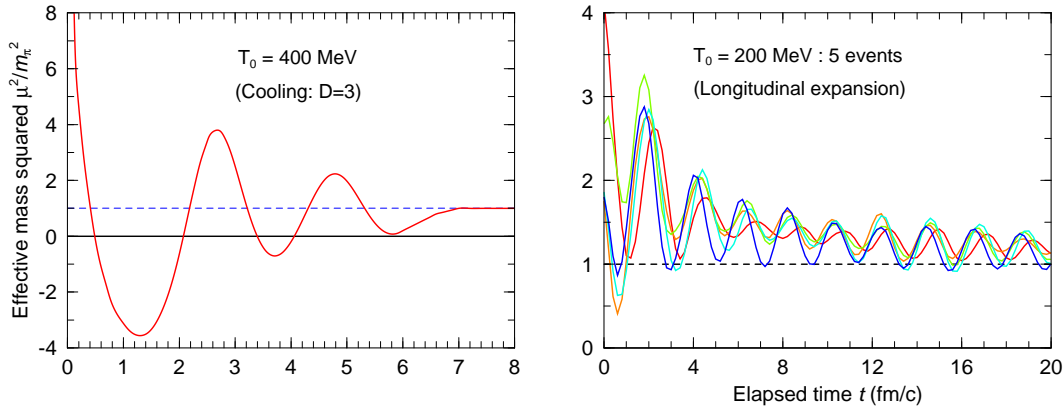}
\vspace*{-1.8cm}
\caption{Initial number of quanta.}
{\small
The time dependence of the square of the effective mass, $\mu^2(t)$,
in the two scenarios considered.
{\em Left panel:}
The quench scenario obtained by first preparing the system
at $T_0=400\ \MeV$ and then cooling with a rate emulating a
three-dimensional scaling expansion;
taken from ref.\ \cite{JR:PRL} with the oscillations suppressed
beyond $t=6\ \fm/c$.
{\em Right panel:}
The moderate scenario in which five initial conditions are sampled
from the thermal ensemble prepared at $T_0=200\ \MeV$
and then evolved up to $t=20\ \fm/c$.
}\label{f:mu2}
\end{figure}

For the quench scenario
we employ the time-dependent effective pion mass
obtained in ref.\ \cite{JR:PRL} by subjecting matter in initial
equilibrium at $T_0=400\ \MeV$ to a Rayleigh cooling with $D$=3.
(which emulates the effect of a three-dimensional scaling expansion
\cite{GM,Lampert}).
%which emulates a three-dimensional scaling expansion of the Bjorken type.
The resulting $\mu^2$ is shown in fig.\ \ref{f:mu2}.
Starting with a high value (reflecting the inital thermal mass),
it rapidly drops into the negative regime,
followed by large but decreasing oscillations
(the amplitudes drop off as $\sim t^{-3/2}$).
For convenience,
the oscillations have been artificially suppressed above $t\approx6\ \fm/c$.

%..............................................................................
%..............................................................................

In the more moderate scenario,
where the systems starts from $T_0=200\ \MeV$,
we have assumed that the system experiences a standard
longitudinal scaling expansion and solved the corresponding
field equation by switching from $(t,z)$ to $(\tau,\eta)$
(see, for example, refs.\ \cite{CooperPRD51,Amado,AHW}).
Since the order parameter is initially quite large (about 60 MeV),
the system never enters the unstable regime and
the effective mass never drops much below the free value.
But the oscillations nevertheless persist for a long time
(their amplitude drops off as $\sim t^{-1/2}$)
and we have truncated the evolution at $t=20\ \fm/c$.
In order to illustrate to effect of event-by-event fluctuations,
we show evolutions that start from five different initial conditions,
each one sampled from the thermal ensemble.\footnote{The
interested reader is referred to ref.\ \cite{JR:PRD}
for a discussion of how the initial state can be sampled
from a thermal ensemble of semi-classical fields.}
Since the number of quanta in each individual mode is stochastic
(governed by a Poisson distribution),
the resulting effective mass also exhibits fluctuations.
Nevertheless, the different evolutions lead to
slowly subsiding and fairly regular oscillations
which have approximately the same frequency.

%----------------------------------------------------------------------------
%\subsection{Enhancement of pions}

With the system initialized as described above,
and with the time dependence of $\mu^2(t)$  given,
it is now straightforward to solve the equations of motion
as described earlier and obtain the number enhancement coefficient $X_N$
for each particular pion mode.
%The result is shown in the left panel of fig.\ \ref{f:400}
%as a function of the magnitude of the pion momentum $p$.
We will discuss the two scenarios in turn.

%............................................................................
\subsection{Quench}
\label{quench}

\begin{figure}[htb]
%\vspace*{-1.0cm}
%\vspace*{-2.9cm}
%		 \insertplot{f400.ps}
\vspace{80mm}
\includegraphics{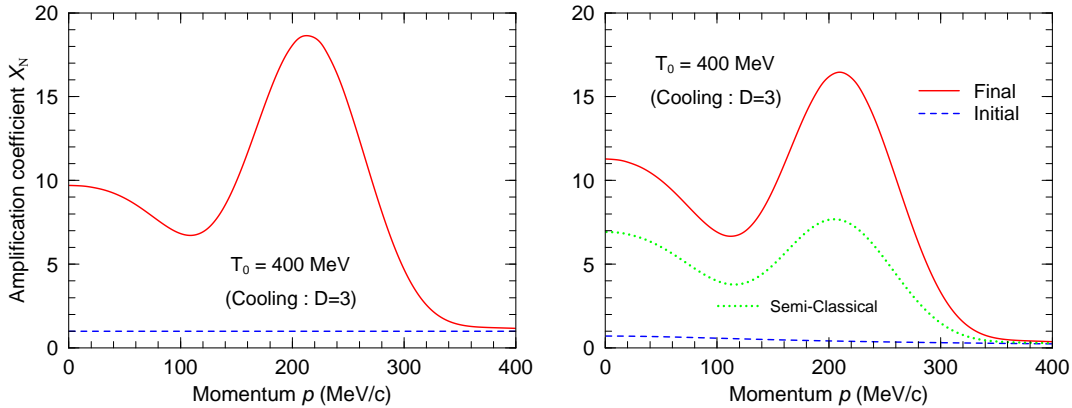}
\vspace*{-1.7cm}
\caption{Result of the quench scenario.}
{\small
{\em Left panel:}
The final value of the number amplification coefficient $X_N$
as a function of the magnitude of the pion momentum,
calculated with the time-dependent mass associated with the quench scenario
(left part of fig.\ \ref{f:mu2}).
{\em Right panel:}
The corresponding final invariant differential pion yield. 
Solid curve: the final yield of free pions;
dashed curve: the number of real quanta in the initial state;
dotted curve: the final yield obtained with a semi-classical treatment.
}\label{f:400}
\end{figure}

The quenched evolution can be regarded as a combination
of the supercritical and oscillating scenarios
considered in sects.\ 5.1 and 5.3, respectively. %\ref{neg} and \ref{osc}
The excursions into the supercritical regime
lead to a significant degree of pion production.
This effect is largest for the softest modes,
since those have the smallest $\omega^2$.
As a result, 
the amplification coefficient $X_N$ grows steadily 
as $p$ is decreased towards zero.
The oscillations in $\mu^2$ lead to the prominent peak near $p=200\ \MeV/c^2$,
reminiscent of the broad peak in fig.\ \ref{f:osc2-X}.
If the oscillations had not been truncated,
this component would have become even more dominant.
The resulting amplification coefficient 
is shown in Fig.~\ref{f:400} (left panel).
It exceeds unity by about one order of magnitude up to $p\approx260\ \MeV$,
at which point it quickly drops to unity.

The final yield of pions can now readily be obtained.
For each value of $p$ we know the initial number of real quanta $\nu_0$
(given by the dashed curve in fig.\ \ref{f:f0}),
and the final number of free pions is then given by
$\nu_f=(\nu_0+\half)X_N-\half$, as discussed earlier.
The resulting spectral profile is shown in the right panel of fig.\ \ref{f:400}.
The initial values $\nu_0$ are about one half
throughout the region affected (see fig.\ \ref{f:f0}),
\ie\ the thermal fluctuations are comparable to the vacuum fluctuations.
The resulting enhancement of the yield
is therefore roughly equal to twice the value of $X_N$,
which amounts to a factor of 20-30 up to $260\ \MeV/c$,
in the particular case presented.

%..............................................................................
%..............................................................................

In a semi-classical treatment of the field theory,
the initial field amplitude does not contain
contributions from vacuum fluctuations.
Consequently,
the final yield would be given by
\beq
\nu_f^{\rm SC}\ =\ \nu_0\ X_N\ <\ (\nu_0+\half)X_N-\half\ =\ \nu_f^{\rm Qu}\ ,
\eeq
as already discussed in Sect.\ 5.4.	%\ref{sc}.
This result is included in fig.\ \ref{f:400}.
If the initial occupation is large, $\nu_0\gg1$,
then the semi-classical result is a good approximation.
However,
in the cases of present interest,
the occupancies are usually at most about one half,
so the semi-classical calculation should be expected
to underestimate the enhancement by at least a factor of two.
Thus, for example, the enhancements obtained in ref.\ \cite{JR:PRL}
with the semi-classical treatment of the linear $\sigma$ model
should be regarded as lower bounds on the expected effect.

Finally,
we wish to express the opinion that the present quench scenario
is probably rather optimistic.
In order for the quench to occur,
approximate chiral symmetry must first be generated
throughout the volume considered;
this is usually thought to occur as a result of the expected
high degree of agitation produced in the early part of the high-energy
collision.
Subsequently,
the system must cool sufficiently rapidly
so as to recover approximately the low-temperature effective potential
before the order parameter has relaxed back to its vacuum value.
This delicate race requires a very rapid expansion.
A variety of calculations suggest that the usual longitudinal
scaling expansion is insufficient \cite{JR:PRL,AHW}
and it may be questionable
whether the onset of the ultimate transverse expansion could occur
before the inherent longitudinal expansion had already caused
the order parameter to grow too large for a supercritical condition to arise.
It therefore appears reasonable to assume that the quench scenario
represents an extreme limit. 

%............................................................................
\subsection{Fizzle}

We now turn to the (probably more likely) scenario,
in which the early dynamics does not achieve approximate restoration
of chiral symmetry.
As a specific example,
we assume that the initial condition corresponds to a temperature of
``only'' $T_0=200\ \MeV$.
Since this is still a bit below the range
at which the order parameter decreases most rapidly
(see for example refs.\ \cite{JR:PRD,Hatsuda} for specific recent calculations
within the linear $\sigma$ model),
the equilibrium value of the order parameter is still fairly high,
$\sigma_0\approx60\ \MeV$,
and even a three-dimensional scaling expansion would not
be able to bring the system into the supercritical region \cite{JR:PRL}.
Therefore,
the primary enhancement mechanism is the relatively moderate oscillations
in the neighborhood of the ground state.

\begin{figure}[htb]
%\vspace*{-1.0cm}
%\vspace*{-2.9cm}
%		 \insertplot{f200.ps}
\vspace{80mm}
\includegraphics{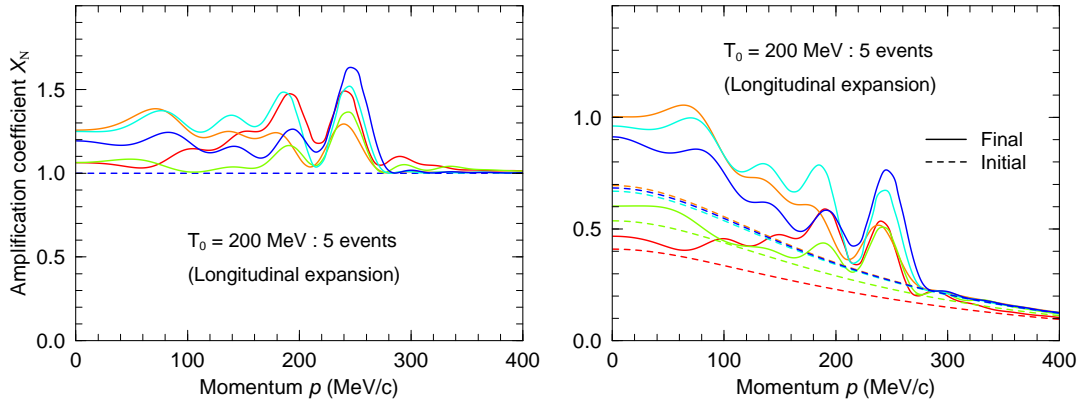}
\vspace*{-1.7cm}
\caption{Result of the fizzle scenario.}
{\small
{\em Left panel:}
The final value of the number amplification coefficient $X_N$
as a function of the magnitude of the pion momentum,
calculated with the time-dependent mass associated with the fizzle scenario
(right part of fig.\ \ref{f:mu2}).
{\em Right panel:}
The corresponding final invariant differential pion yield. 
Solid curve: the final yield of free pions;
dashed curve: the number of real quanta in the initial state.
}
\label{f:200}
\end{figure}

In the left panel of fig.\ \ref{f:200} is shown the resulting values of the
amplification coefficient $X_N$ as a function of the momentum $p$,
for each of the five histories based on the initial conditions
that were sampled from the thermal ensemble of quasiparticles
at the given $T_0$.
Because of the differences in the microscopic composition of the
five initial states, the detailed evolutions will differ somewhat and,
as a consequence, we will obtain an entire ensemble of resulting $X_N(p)$.
However, although they differ in detail,
the various histories tend to yield qualitatively similar results,
as is evident from the display.

The result can be characterized as a broad and modest increase of $X_N$
by about 20\% up to about $p\approx240\ \MeV/c$
plus a relatively narrow peak centered at about that momentum,
followed by an abrupt drop-off at higher momenta.
The peak is positioned where we would expect:
It follows from the discussion in Sect.\ 5.3 %\ref{osc}
that the modes most affected
by the quasi-regular oscillations of the mass
are those with a frequency just below half the $\sigma$ mass
(the calculations use $m_\sigma=600\ \MeV/c^2$).
Since the effective pion mass is about $\mu_{\rm ave}\approx 150\ \MeV$
through the later part of the evolution,
the frequency associated with the most affected mode is about
$\omega\approx285\ \MeV$ which is indeed slightly less than $\half m_\sigma$.
We note that in the present calculation
the oscillations in the effective mass were artificially eliminated
after $t_{\rm max}=20\ \fm/c$,
even though the system would keep ringing for a longer time.
Naturally,
if $t_{\rm max}$ were increased then the resulting amplification
coefficients would be larger.

%..............................................................................
%..............................................................................

The effect of the enhancement on the pion yield is finally shown
in the right panel of fig.\ \ref{f:200} for all five events,
using the same procedure as in the quench scenario.
Based on these calculations one would expect a rather broad enhancement
of the pion yield at low momenta momenta, $p<250\ \MeV/c$,
corresponding to kinetic energies below about $150\ \MeV$.
This enhancement is fairly modest,
amounting to perhaps 50\%,
but might be visible as an excess relative to a thermal extrapolation
based on the yield at higher energies.
In addition, there appears a fairly robust peak near
the upper edge of the enhanced region.
Its magnitude is probably difficult to predict reliably,
since it is sensitive to the amplitude and persistence
of the induced oscillations in the chiral order parameter.
Furthermore,
although the peak location can be pin-pointed rather well within
the model calculation,
it should be kept in mind that the determining quantity
is the $\sigma$ mass which is not well established.
In fact, the $\sigma$ field is introduced as merely a simplistic representation
of the observed rather broad enhancement
of the elementary isoscalar cross section.
Consequently, it might be too naive to expect well localized peak
to appear in the observed spectra.

%============================================================================
\section{Concluding remarks}

In the present work,
we have discussed the quantum field treatment of free particles
exposed to an arbitrary time-dependent external field that may
become supercritical so that spontaneous pair creation occurs.
Though of general nature,
this problem is particularly relevant for current efforts
to explore the possibility for forming so-called disoriented chiral condensates
in high-energy nuclear collisions,
because the key effect is expected to be the amplification of soft pionic
modes by either a transient supercritical field 
or the eventual oscillatory relaxation of the order parameter
near its vacuum value.

%Begin addition:
For the treatment of the time-dependent quantum-field problem,
we adapt a method introduced by Combescure \cite{Combescure};
this method was also employed in Ref.\ \cite{Amado}.
It has some resemblence with the the treatment
of the corresponding quantum-mechanical problem
(where the goal is to obtain the wave function for a forced oscillator)
introduced by Husimi \cite{Husimi}.
%End of addition.
We have developed the general formalism and illustrated it both
for instructive idealized cases and 
cases of direct DCC relevance.
By divorcing the specific initial state from the behavior
of the evolution operator,
the method makes it easy to treat the problem
for any given time dependence of the frequency, $\omega^2(t)$.
In particular, the expected final number of quanta in a mode, $\nu_f$,
is simply given in terms of the initial occupancy $\nu_0$
by means of an amplification coefficient $X_N$ that depends on $\omega^2(t)$
but not on $\nu_0$.
In addition to being convenient,
this form is also instructive,
as it brings out the fact that the vacuum fluctuations
and thermal fluctuations contribute in a democratic fashion,
both being subject to the same amplification by the factor $X_N$.
In particular,
the result makes it easy to see how spontaneous particle production
is possible even when starting in the vacuum state.

As a supplement to the quantum-field developments,
we also showed how the same key results can be obtained
entirely within classical mechanics by performing a suitable
analysis of how the results depend on the initial phase of the state.
Although we suspect that this novel method is in fact mathematically
equivalent to the quantal treatment,
we have only demonstrated the correspondence
by means of numerical calculations.
The fact that processes that are inherently quantal in nature,
such as spontaneous production in supercritical fields,
can be treated accurately within the confines of classical physics
may prove to be practically useful,
since it is significantly easier to treat the field dynamics classically.

Moreover,
the formal discussion also makes it possible to understand the
quantitative limitations of a semi-classical treatment of the field equation.
This is important,
since by far most of the dynamical calculations
in the context of disoriented chiral condensates
have in fact been carried out classically.
The developed classical method for obtaining the amplification coefficients
may possibly be generalized and brought to bear on the evolution
of the entire field in order to extract more reliable
information from essentially classical calculations.

Finally,
employing time-dependent effective masses obtained from dynamical
calculations within the linear $\sigma$ model,
we have illustrated the degree of amplification
that can occur in possible collision scenarios.
In particular,
a significant enhancement of certain modes is expected to occur
as a consequence of the quasi-regular behavior of the chiral condensate,
as it relaxes after its initial agitation far away from its vacuum value.
In a rather extreme scenario,
where the cooling rate emulates a scaling expansion in three dimensions
and a transient quench is generated,
the soft pion modes are enhanced by well over an order of magnitude.
A more cautious scenario,
where neither the initial chiral restoration
nor the subsequent expansion is as extreme,
yields a modest but persistent enhancement of the soft modes
as well as a larger preferential enhancement at a finite pion energy
related to the mass of the effective $\sigma$ meson.
This latter effect is a parametric resonance phenomenon
whose appearance in the context of disoriented chiral condensates
has already been recognized by several groups
(see, for example, Refs.\ \cite{BoyanovskyPRD51a,MM,ee}).

%Begin addition:
The presented applications serve primarily as illustrations only and
we have limited the discussion to the enhancement of the one-particle spectra.
It has been suggested that the time dependence of the effective mass
may produce a substantial signal in the two-particle correlations \cite{AC96}
and recent studies have explored the possibility of using this effect
as a DCC signal \cite{HM98}.
The present treatment may provide a helpful tool for further explorations
of this interesting prospect.
%End of addition.

Once the time dependence of the effective mass has been calculated,
in a suitable model,
the present formalism makes it possible to calculate the resulting
effect on the pion dynamics with due account taken of the complications
arising from the inherent quantum nature of the system.
In this regard,
we expect that the present work will be of some practical utility.
In particular,
it might be possible to use it as a means
for making a perturbative estimate of the quantum enhancement. 
Of course,
it would be desirable to extend the treatment to the more general case
where the effective mass is endowed with a spatial variation as well;
this problem presents an interesting future challenge.

%==============================================================================
~\\ \noindent
Helpful and stimulating discussions with Andrei Krzywicki
are gratefully acknowledged.
This work was supported by the Director, Office of Energy Research,
Office of High Energy and Nuclear Physics,
Nuclear Physics Division of the U.S. Department of Energy
under Contract No.\ DE-AC03-76SF00098.

%==============================================================================
\section*{Appendix: Coherent states}
\label{coherent}
We here recall some of the features of coherent states
that may be especially useful in the context of the present study.

One would ordinarily like to associate
a free field with a specific coherent state,
\beq\label{chi}
|\chi\rangle\ \equiv\ 
\rme^{\sum_\cappa[\chi_\cappa^{} \alpha_\cappa^\dagger
-\chi_\cappa^* \alpha_\cappa^{}]} |0\rangle\ =\
{\cal N}_\chi\rme^{\sum_\cappa\chi_\cappa^{} \alpha_\cappa^\dagger}|0\rangle\ ,
\eeq
where ${\cal N}_\chi=\exp(-{1\over2}\sum_\cappa|\chi_\cappa|^2)$.
Thus $|\chi\rangle$ is normalized to unity,
\beq
\langle\chi|\chi'\rangle\ =\
{\cal N}_\chi {\cal N}_{\chi'}\ 
\rme^{\sum_\cappa \chi_\cappa^* \chi_\cappa'}\ .
\eeq
Furthermore, $\alpha_\cappa |\chi\rangle=\chi_\cappa|\chi\rangle$ and
$\langle\chi|\alpha_\cappa^\dagger \alpha_{\cappa'}^{}|\chi'\rangle
=\chi_\cappa^*\chi_{\cappa'}'\langle\chi|\chi'\rangle$.
For free motion the time evolution of $|\chi\rangle$
is governed by $\del_t{\chi}_\cappa=-i\zero{\omega}_\cappa\chi_\cappa$.
We also note that
\beqar
\langle\chi|\hat{q}_\cappa|\chi\rangle &=&
{1\over\sqrt{2\zero{\omega}_\kappa}}
\langle\chi|\alpha_\cappa^{}+\alpha_\cappa^\dagger|\chi\rangle\ =\
\sqrt{2\over\zero{\omega}_\kappa}\ {\rm Re}\chi_\kappa\ ,\\
\langle\chi|\hat{p}_\cappa|\chi\rangle &=&
i{\sqrt{\zero{\omega}_\kappa\over2}}
\langle\chi|\alpha_\cappa^\dagger-\alpha_\cappa^{}|\chi\rangle\ =\
\sqrt{2\zero{\omega}_\kappa}\ {\rm Im}\chi_\kappa\ ,
\eeqar
so that
$\chi_\cappa=\sqrt{{\omega}_\kappa/2}q_\cappa
+i/\sqrt{2{\omega}_\kappa}p_\cappa$.
Thus, if we write $\chi_\cappa=\sqrt{\nu_\cappa}\exp(-i\eta_\cappa)$,
we have
\beq
q_\cappa=\sqrt{2\nu_\cappa/\zero{\omega}_\kappa}\cos\eta_\cappa\ ,\
p_\cappa=-\sqrt{2\nu_\cappa\zero{\omega}_\kappa}\sin\eta_\cappa\ .
\eeq

We also note that when $\omega^2$ is constant in time,
then the coherent state (\ref{chi}) is a solution to the equation of motion,
$i\del_t|t\rangle=\hat{H}|t\rangle$,
where $|t\rangle\equiv\exp(\hat{\Chi})|0\rangle$ with
\beq
\hat{\Chi}(t)\ \equiv\ \chi\alpha^\dagger-\chi^*\alpha\
=\ i(p\hat{q}-q\hat{p})\ .
\eeq
We have here used the relations (\ref{qkhat}-\ref{pkhat})
to introduce the canonical operators $\hat q$ and $\hat p$
(which are independent of time).
The new parameters (which are real)
are related to the original (complex) parameter $\chi(t)$ by
\beq
\chi(t) =\ \sqrt{\omega_0\over2}\ q(t)+{i\over\sqrt{2\omega_0}}\ p(t)\ 
=\ \langle\alpha\rangle
\eeq
and they represent the expectation values
of the canonical operators,
$q(t)=\langle t|\hat{q}|t\rangle$ and $p(t)=\langle t|\hat{p}|t\rangle$,
as the notation would suggest.
We then note that the evolution of the state is given by
$|t\rangle=\exp(-i\hat{H}t)|t_0\rangle$ when $\hat{H}$ is time-independent.
It possible to show that the resulting state is still of coherent form
and that the parameters $q$ and $p$ satisfy Hamilton's equations,
\beq\label{qp}
\dot{q}\ =\ p\ ,\,\,\,\ \dot{p}\ =\ -\omega^2 q\ .
\eeq
Thus, in summary,
when the Hamiltonian is of harmonic form,
$H=\hat{p}^2/2+\omega^2\hat{q}^2/2$,
then the coherent state $|t\rangle\equiv\exp(i(p\hat{q}-q\hat{p})|0\rangle$
is a solution to the equation of motion provided the parameters
satisfy the corresponding Hamilton equations,
$\dot{q}=p$ and $\dot{p}=-\omega^2q$.
This result does not depend on the sign of $\omega^2$
but relies only on its constancy.

The above coherent state (\ref{chi}) is written in terms of the
operators $\alpha_\cappa$ associated with the diagonal basis.
The state can equally well be written in terms of the original operators
$a_\k$ which annihilate quanta having a definite momentum $\k$,
\beq
|\chi\rangle\ \equiv\ 
\rme^{\sum_\cappa[\chi_\cappa^{} \alpha_\cappa^\dagger
-\chi_\cappa^* \alpha_\cappa^{}]} |0\rangle\ =\
\rme^{\sum_\k[\chi_\k^{} a_\k^\dagger
-\chi_\k^* a_\k^{}]} |0\rangle\ .
\eeq
The parameters $\chi_\k$ are easily obtained.
With $\cappa>0$ we find,
\beq
\chi_{\k=\cappa}={1\over\sqrt{2}}[\chi_{\cappa}-i\chi_{-\cappa}]
\,\,\ ,\,\,
\chi_{\k=-\cappa}={1\over\sqrt{2}}[\chi_{\cappa}+i\chi_{-\cappa}]\ .
\eeq

A coherent state of effective (or `squeezed') quanta
(which is what would ordinarily be used as an initial state),
is also a coherent state in terms of the original free representation,
\beq\label{chi0}
|\tilde{\chi}\rangle\
\equiv\ \rme^{\tilde{\chi}\beta^\dagger-\tilde{\chi}^*\beta}|0\rangle
=\ \rme^{\chi\alpha^\dagger-\chi^*\alpha}|0\rangle\ \equiv\ |\chi\rangle\ ,
\eeq
where $\chi_k$ denote the coefficients pertaining to the free representation,
\beq
\chi\ =\ {\tilde{\chi}+\gamma\tilde{\chi}^* \over \sqrt{1-g^2}}\
=\  \left({\omega_0\over\omega}\right)^{1\over2}{\rm Re}\tilde{\chi}\
+\ i\left({\omega\over\omega_0}\right)^{1\over2}{\rm Im}\tilde{\chi}\ ,
\eeq
and we have used that
$\tilde{\chi}\beta^\dagger-\tilde{\chi}^*\beta
= \chi\alpha^\dagger-\chi^*\alpha$.
Moreover,
it has been assumed that the effective frequency is positive, $\omega>0$,
as will be hte case if the quasiparticle modes are in thermal equilibrium.
This relationship can be rewritten on a simple democratic form,
\beq
\tilde{\chi}_\cappa\ =\ 
\sqrt{\omega\over2}\ q_\cappa+{i\over\sqrt{2\omega}}\ p_\cappa\ 
\Longleftrightarrow\
\chi_\cappa\ =\ 
\sqrt{\omega_0\over2}\ q_\cappa+{i\over\sqrt{2\omega_0}}\ p_\cappa\ , 
\eeq
where $q_\cappa$ and $p_\cappa$ are the expextation values
of the canonical operators in the diagonal representation,
as given by the (real) coefficients in the trigonometric expansions
(\ref{qk}-\ref{pk}).

%=============================================================================

\vfill\eject
%==============================================================================
                        \end{document}